\documentclass[aps,prb,showpacs,reprint,superscriptaddress]{revtex4-1}

\usepackage[utf8]{inputenc}
\usepackage[pdftex]{graphicx}
\usepackage{color}
\usepackage{float} 
\usepackage{amssymb}
\usepackage{amsmath}
\usepackage{ragged2e}

\begin{document}

\title{Spin dynamics in the antiferromagnetic phases of the\\ Dirac metals $A$MnBi$_2$ ($A=$ Sr, Ca)}

\author{M. C. Rahn}
\email[]{marein.rahn@physics.ox.ac.uk}
\affiliation{Clarendon Laboratory, Department of Physics, University of Oxford, Oxford, OX1 3PU, United Kingdom}

\author{A. J. Princep}
\affiliation{Clarendon Laboratory, Department of Physics, University of Oxford, Oxford, OX1 3PU, United Kingdom}

\author{A. Piovano}
\affiliation{Institut Laue-Langevin, 6 rue Jules Horowitz, 38042 Grenoble Cedex 9, France}

\author{J. Kulda}
\affiliation{Institut Laue-Langevin, 6 rue Jules Horowitz, 38042 Grenoble Cedex 9, France}

\author{Y. F. Guo}
\affiliation{School of Physical Science and Technology, ShanghaiTech University, Shanghai 201210, China}
\affiliation{CAS Center for Excellence in Superconducting Electronics (CENSE), Shanghai 200050, China}

\author{Y. G. Shi}
\affiliation{Beijing National Laboratory for Condensed Matter Physics, Institute of Physics, Chinese Academy of Sciences, Beijing 100190, China}

\author{A. T. Boothroyd}
\email[]{a.boothroyd@physics.ox.ac.uk}
\affiliation{Clarendon Laboratory, Department of Physics, University of Oxford, Oxford, OX1 3PU, United Kingdom}

\date{\today}

\begin{abstract}
The square Bi layers in $A$MnBi$_2$ ($A=$ Sr, Ca) host Dirac fermions which coexist with antiferromagnetic order on the Mn sublattice below $T_{\rm N} = 290$\,K (Sr) and 270\,K (Ca). We have measured the spin-wave dispersion in these materials by triple-axis neutron spectroscopy. The spectra show pronounced spin gaps of $10.2(2)\,$meV (Sr) and $8.3(8)\,$meV (Ca) and extend to a maximum energy transfer of 61--63\,meV. The observed spectra can be accurately reproduced by linear spin-wave theory from an Heisenberg effective spin Hamiltonian. Detailed global fits of the full magnon dispersion are used to determine the in-plane and inter-layer exchange parameters as well as on the magnetocrystalline anisotropy constant. To within experimental error we find no evidence that the magnetic dynamics are influenced by the Dirac fermions.
\end{abstract}


\maketitle

\section{Introduction}

Following the observation of the topological properties of electrons on the honeycomb layers of graphene, the universal characteristics of massless dispersing low-energy quasiparticles have been realized across a variety of condensed matter systems. The ternary bismuthides \textit{A}MnBi$_2$ (\textit{A}$\,=\,$ Ca, Sr) \cite{Cordier1977,Brechtel1980,Shim2009,Wang2011} are a recent addition to this family of so-called Dirac materials. The Bi square layers of \textit{A}MnBi$_2$ have been found to show the same unusual transport characteristics as graphene or topological insulators\cite{Ran2009,Richard2010,Morinari2010,Vafek2014}. Due to the suppression of backscattering processes, the electronic and thermal conductivity are enhanced, and the large separation of Landau levels produces a large linear magnetoresistance. Indeed, angle-resolved photoemission spectroscopy (ARPES) has provided direct evidence of the linear band crossings in both SrMnBi$_2$ and CaMnBi$_2$ \cite{Park2011,Feng2014}, with a highly anisotropic Dirac cone.

Among other Dirac materials, these bismuthides attract special interest because their Dirac fermions may couple to transition-metal states, promising an indirect experimental handle to tune the topological bands. Below $T_{\rm N}^{\text{Sr}}\simeq 290\,$K and $T_{\rm N}^{\text{Ca}}\simeq 270\,$K, the large divalent Mn ($3d^5, S=5/2$) magnetic moments of magnitude $\approx 3.7\,\mu_{\text{B}}$ in these materials align parallel to the $c$-axis and form antiferromagnetic structures\cite{Guo2014}. The two compounds were found to differ in the sign of their interlayer coupling, resulting in ferro- and antiferromagnetic stacking of N\'{e}el-ordered layers in CaMnBi$_2$ and SrMnBi$_2$, respectively \cite{Guo2014}. An interpretation based on first principles calculations suggests that in the ferromagnetically stacked case (CaMnBi$_2$), the Dirac bands may provide an itinerant inter-layer exchange path and thus directly couple to the magnetic ground state\citep{Guo2014}. This appeared to be supported by a weak resistivity anomaly observed at $T_N$ in CaMnBi$_2$, but not in SrMnBi$_2$ \cite{Guo2014}. Earlier transport studies, however, had not registered such an anomaly in either SrMnBi$_2$ (Ref.~\onlinecite{Wang2011}) or CaMnBi$_2$ (Refs.~\onlinecite{Wang2012,He2012}).

In metallic magnets a coupling between the ordered magnetic moments and conduction electron states can reveal itself in the magnetic excitation spectrum. For example, there can be damping due to spin-wave decay into the Stoner continuum, anomalies in the magnon dispersion due to modifications of the exchange interactions by conduction electron states, or gap formation due to an additional Kondo energy scale.

Here we report on a single-crystal neutron inelastic scattering study of SrMnBi$_2$ and CaMnBi$_2$ in the magnetically ordered state. Our analysis shows that the magnon spectrum in both materials can be accurately reproduced from a Heisenberg model describing a local-moment, quasi-two-dimensional (2D) antiferromagnet. The model includes nearest- and next-nearest-neighbor in-plane exchange interactions and a weak inter-layer exchange interaction, together with an easy-axis anisotropy. We did not find any anomalies that would suggest significant coupling between the magnons and conduction electron states. The interlayer coupling is smaller than found in the reference compound BaMn$_2$Bi$_2$, consistent with the larger separation of the Mn spins along the $c$-axis in $A$MnBi$_2$.

\begin{figure}
\includegraphics[width=0.85\columnwidth,trim= 0pt 0pt 0pt 0pt, clip]{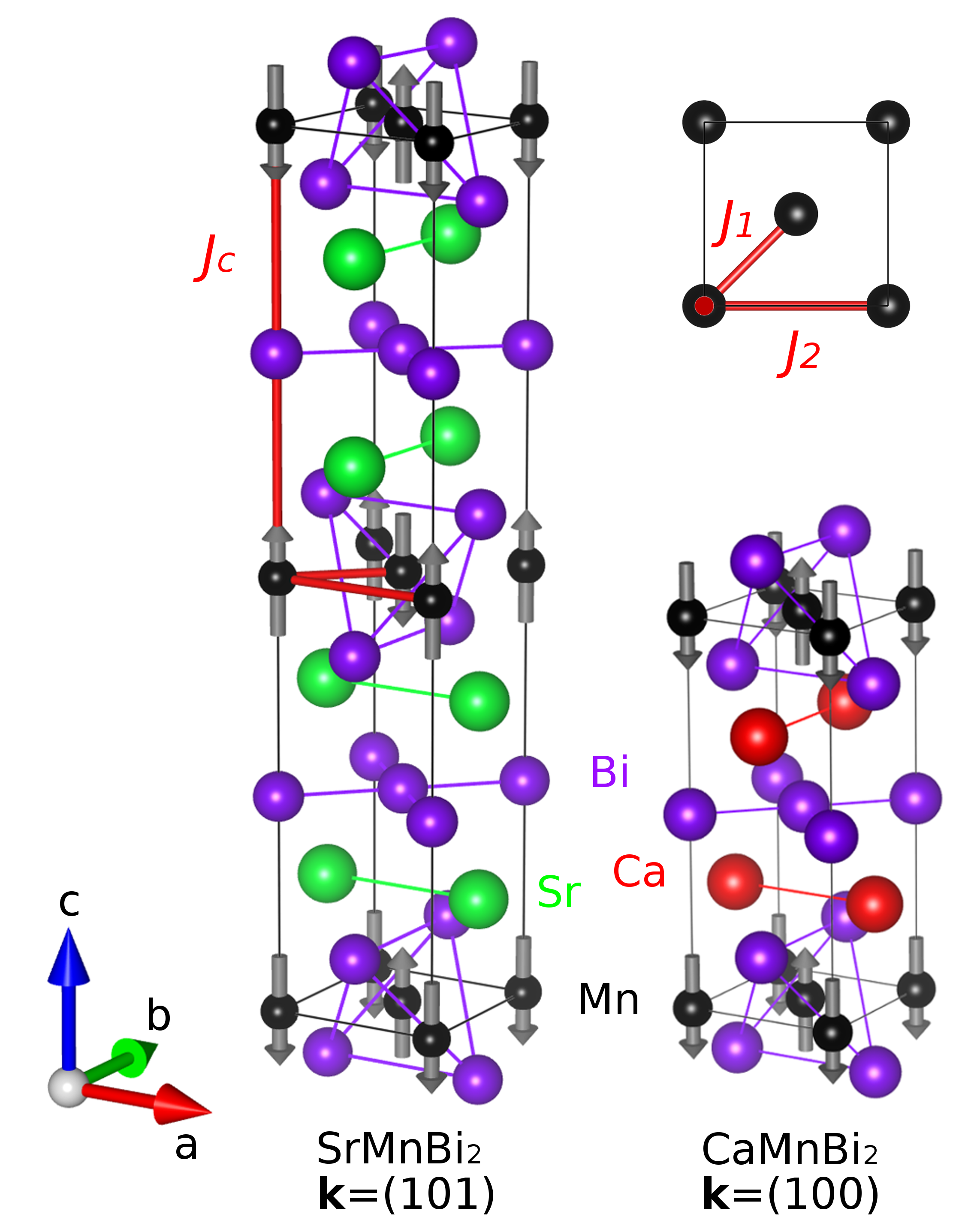}%
\caption{\label{fig1} (color online). The crystal and magnetic structures of SrMnBi$_2$ ($a\approx 4.58\,$\AA, $c\approx 23.14\,$\AA, magnetic space group $I4'/m'm'm$) and CaMnBi$_2$ ($a\approx 4.50\,$\AA, $c\approx 11.07\,$\AA, magnetic space group $P4'/n'm'm$)\citep{Guo2014}. The magnetic propagation vectors $\mathbf{k}$ indicated in this figure describe the magnetic structures $\mathbf{m}(\mathbf{r}_j)$ (in lattice coordinates $\mathbf{r}_j$) by the relation $\mathbf{m}(\mathbf{r}_j) =\mathbf{m}(0)\,\exp(2\pi {\rm i}\,\mathbf{k}\cdot\mathbf{r}_j)$, where $\mathbf{m}(0)$ is the magnetic moment at an arbitrary origin located on a Mn site. For clarity, the origin of the SrMnBi$_2$ unit cell has been shifted by $(\frac12,0,\frac14)$ relative to the conventional cell. The exchange paths $J_1$, $J_2$ and $J_c$ are indicated by red lines.}

\end{figure}

\section{Experimental details}

The preparation and characterisation of the single crystals used in the experiments has been reported previously\cite{Guo2014}. Polycrystalline {\it A}MnBi$_2$ was first synthesized by solid-state reaction of the elements. Single crystals were then grown from self-flux in an alumina crucible. Electron-probe microanalysis confirmed near-ideal stoichiometry, with a small ($\approx 2\%$) Bi deficiency in the Sr compound (for details, see Ref.~\onlinecite{Guo2014}). Laboratory x-ray diffraction measurements confirmed the tetragonal crystal structures reported previously,\cite{Cordier1977,Brechtel1980} space groups $I4/mmm$ (SrMnBi$_2$) and $P4/nmm$ (CaMnBi$_2$), see Fig.~\ref{fig1}. Magnetization measurements on the batch of crystals used here were consistent with previous studies (see supplemental material\cite{supplemental}).

Neutron inelastic measurements were performed at the Institut Laue--Langevin on the triple-axis neutron spectrometer IN8 (Ref.~\onlinecite{Hiess2006}) with the FlatCone detector.\cite{Kempa2006} By keeping the outgoing energy fixed and recording rocking scans at various incident energies, this setup allows an efficient collection of constant energy-transfer maps covering a wide range of reciprocal space. The FlatCone array of analyzer crystals and helium tube detectors consists of 31 channels spaced by 2.5\textordmasculine, thus covering a 75\textordmasculine\, range of scattering angle. Throughout the study, the FlatCone was used with its Si $(111)$ analyzer crystals selecting a fixed outgoing wavevector of $k_{\rm f}=3\,$\AA\, ($E_{\rm f}=18.6\,$meV). For energy transfers below and above 40\,meV (incoming energies $E_{\rm i} \gtrless 58.6\,$meV), the double-focusing Si $(111)$ and pyrolitic graphite (002) monochromators were used, respectively. In four separate experiments, the scattering from the SrMnBi$_2$ and CaMnBi$_2$ single crystals (of mass 3.3\,g and 1.6\,g, respectively) was investigated in the $(HK0)$ ($ab$ orientation) and $(H0L)$ ($ac$ orientation) scattering planes. Throughout this paper we give wavevectors in reciprocal lattice units (r.l.u.) ${\bf q} = (H,K,L) \equiv (H\times 2\pi/a, K\times 2\pi/b, L\times 2\pi/c)$.
The samples were mounted in a standard top-loading liquid helium cryostat. All spectra were recorded at a sample temperature of approximately $5\,$K.

\begin{figure}
\includegraphics[width=0.87\columnwidth,trim= 0pt 0pt 0pt 0pt, clip]{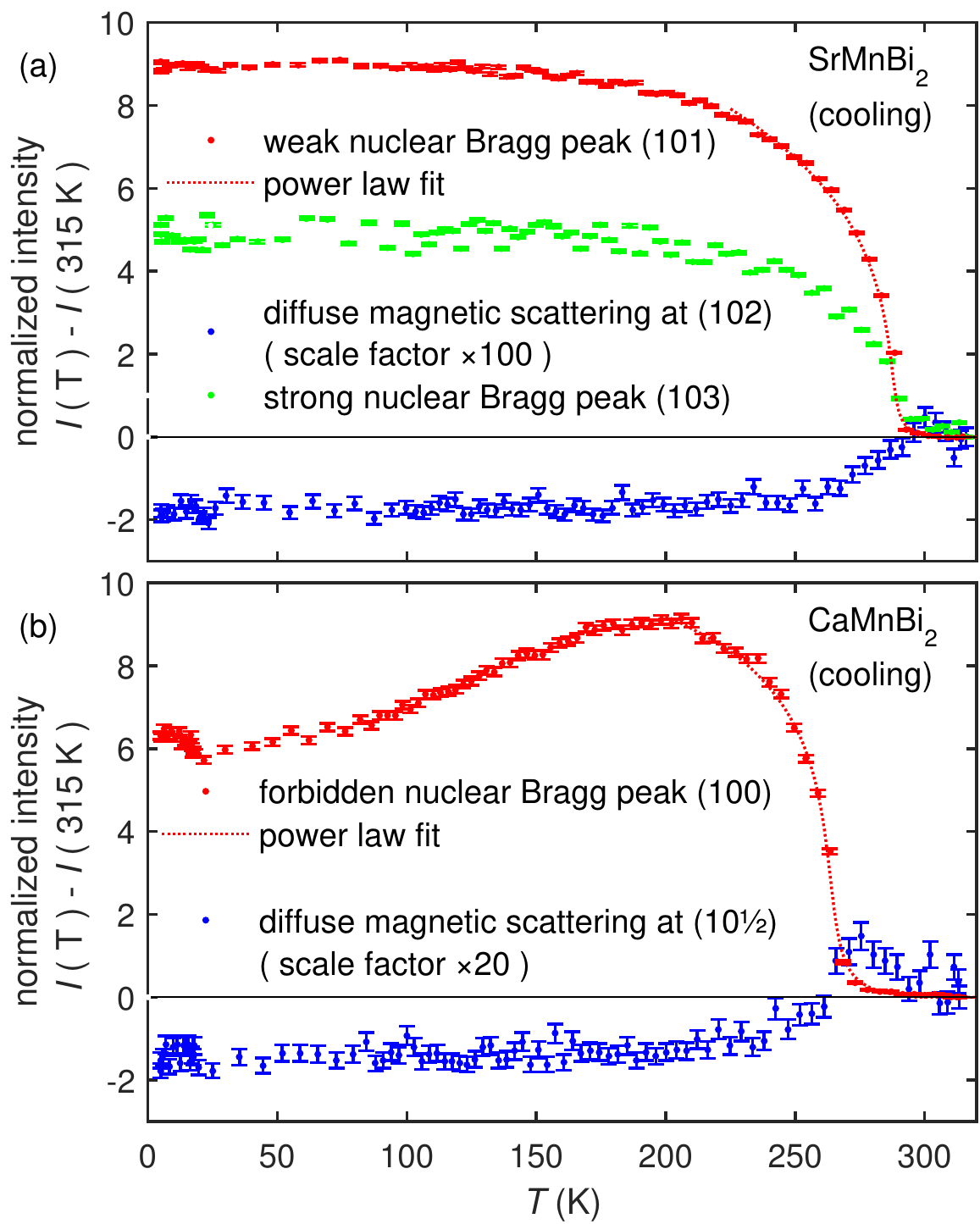}%
\caption{\label{fig2} (color online). Temperature dependence of the difference intensity $I(T)-I(315\,{\rm K})$ at selected wavevectors (see legend), recorded while cooling (a) SrMnBi$_2$ and (b) CaMnBi$_2$. Power law fits to the Bragg peaks yield transition temperatures of $T_{\rm N}^{\text{Sr}}=287(5)\,$K and $T_{\rm N}^{\text{Ca}}=264(2)\,$K. Above $T_{\rm N}$, incipient in-plane correlations contribute diffuse rods of magnetic scattering along $(10L)$. These fluctuations are enhanced towards $T_{\rm N}$ (critical scattering) and then freeze out with the onset of inter-plane order (blue symbols). The decrease in intensity of the (100) reflection of CaMnBi$_2$ below 200$\,$K is not consistent with previous data and should be disregarded (see main text).}  
\end{figure}

\begin{figure*}
\includegraphics[width=1.9\columnwidth,trim= 0pt 0pt 0pt 0pt, clip]{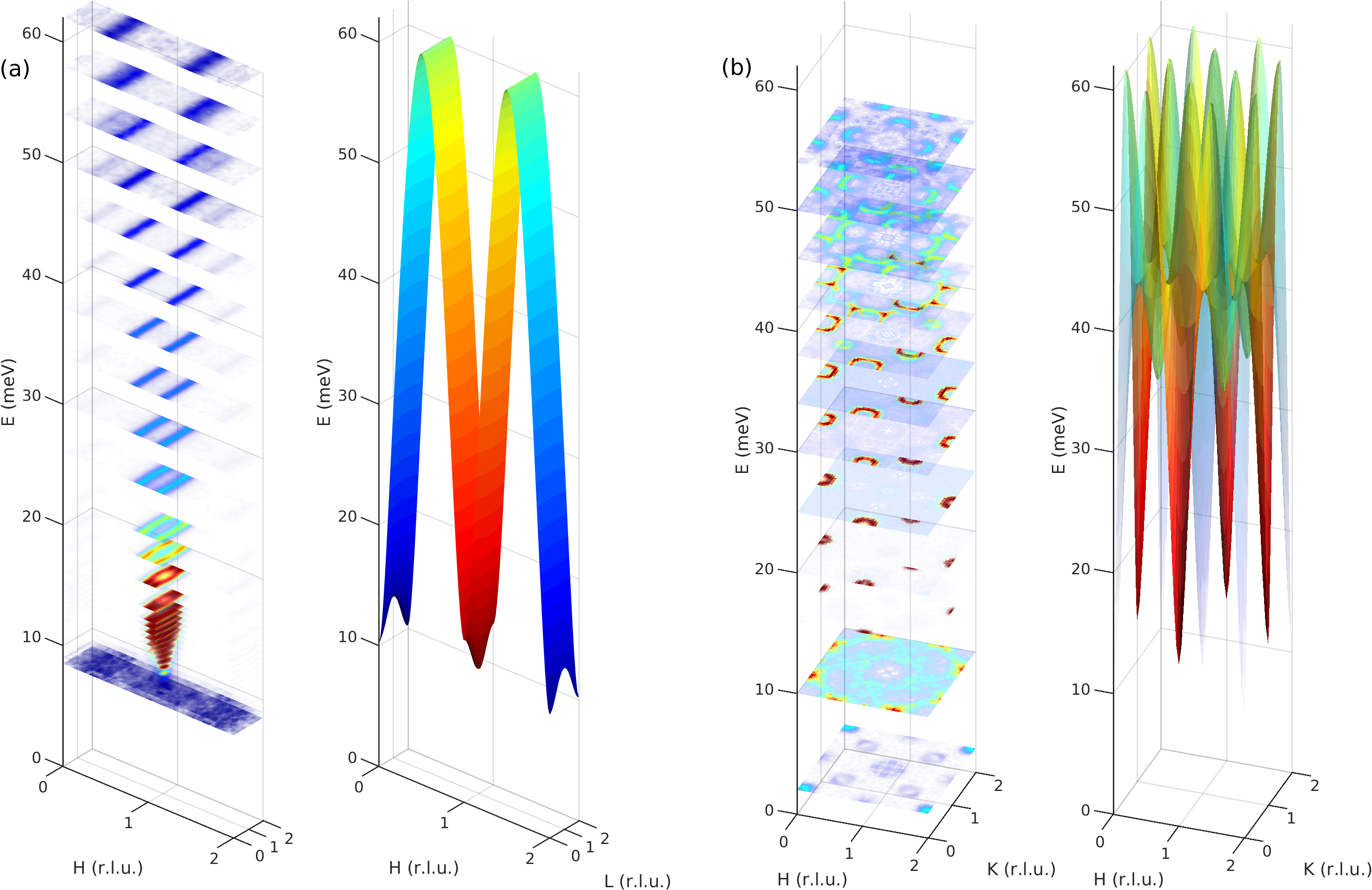}%
\caption{\label{fig3} (color online). Magnon spectrum of SrMnBi$_2$ in the (a) $(H0L)$ and (b) $(HK0)$ planes in reciprocal space. The data are illustrated by a stacking plot of constant-energy slices (left panels), and the best-fit spin-wave model is represented by the corresponding dispersion surface (right panels). A more quantitative comparison between data and simulation is provided in the Supplemental Material.\cite{supplemental} }
\end{figure*}

\begin{figure*}
\includegraphics[width=1.9\columnwidth,trim= 0pt 0pt 0pt 0pt, clip]{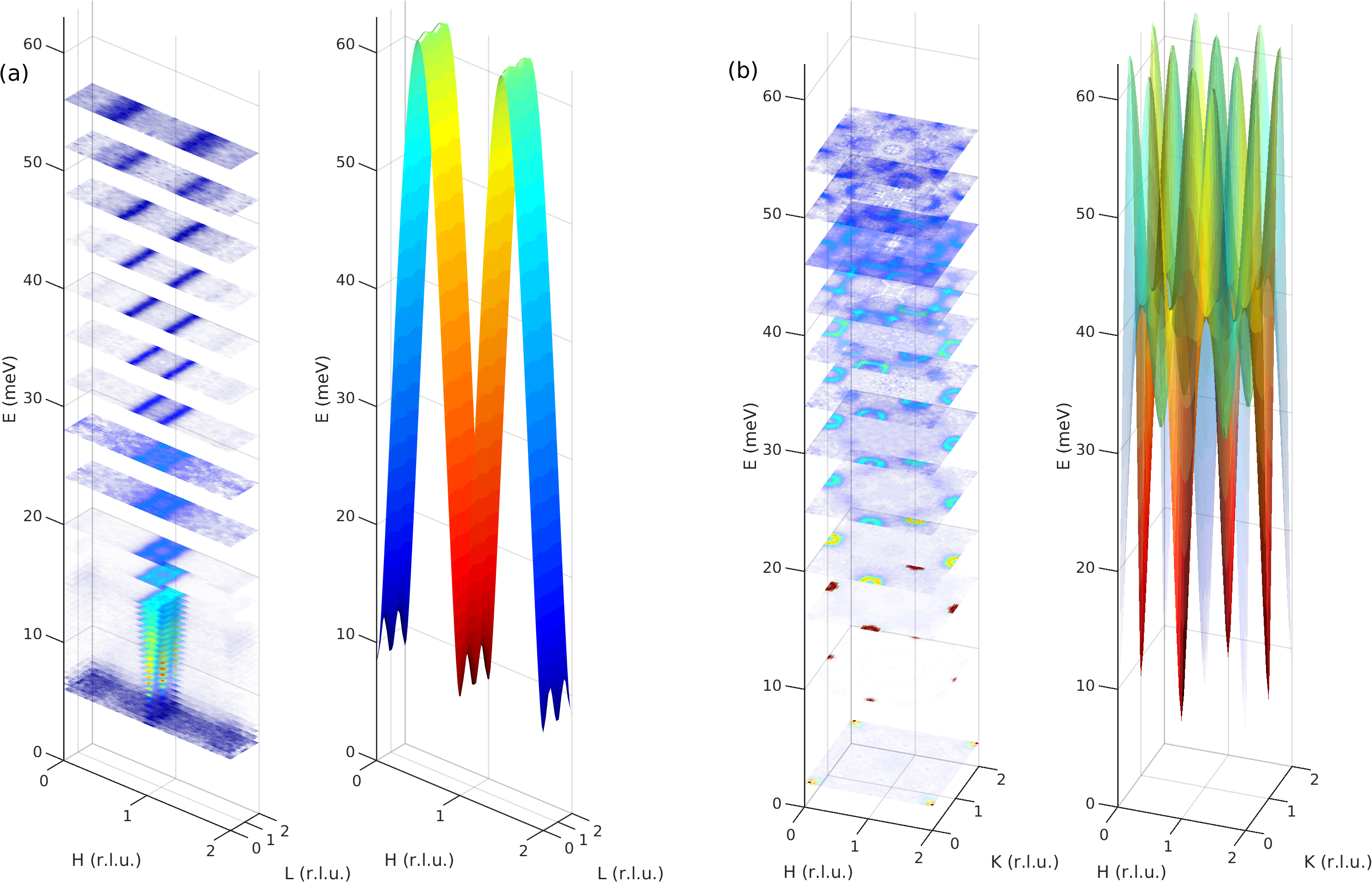}%
\caption{\label{fig4} (color online). Magnon spectrum of CaMnBi$_2$ presented in the same way as in Fig.~\ref{fig3}. }
\end{figure*}

\begin{figure}
\includegraphics[width=0.95\columnwidth,trim= 0pt 0pt 0pt 0pt, clip]{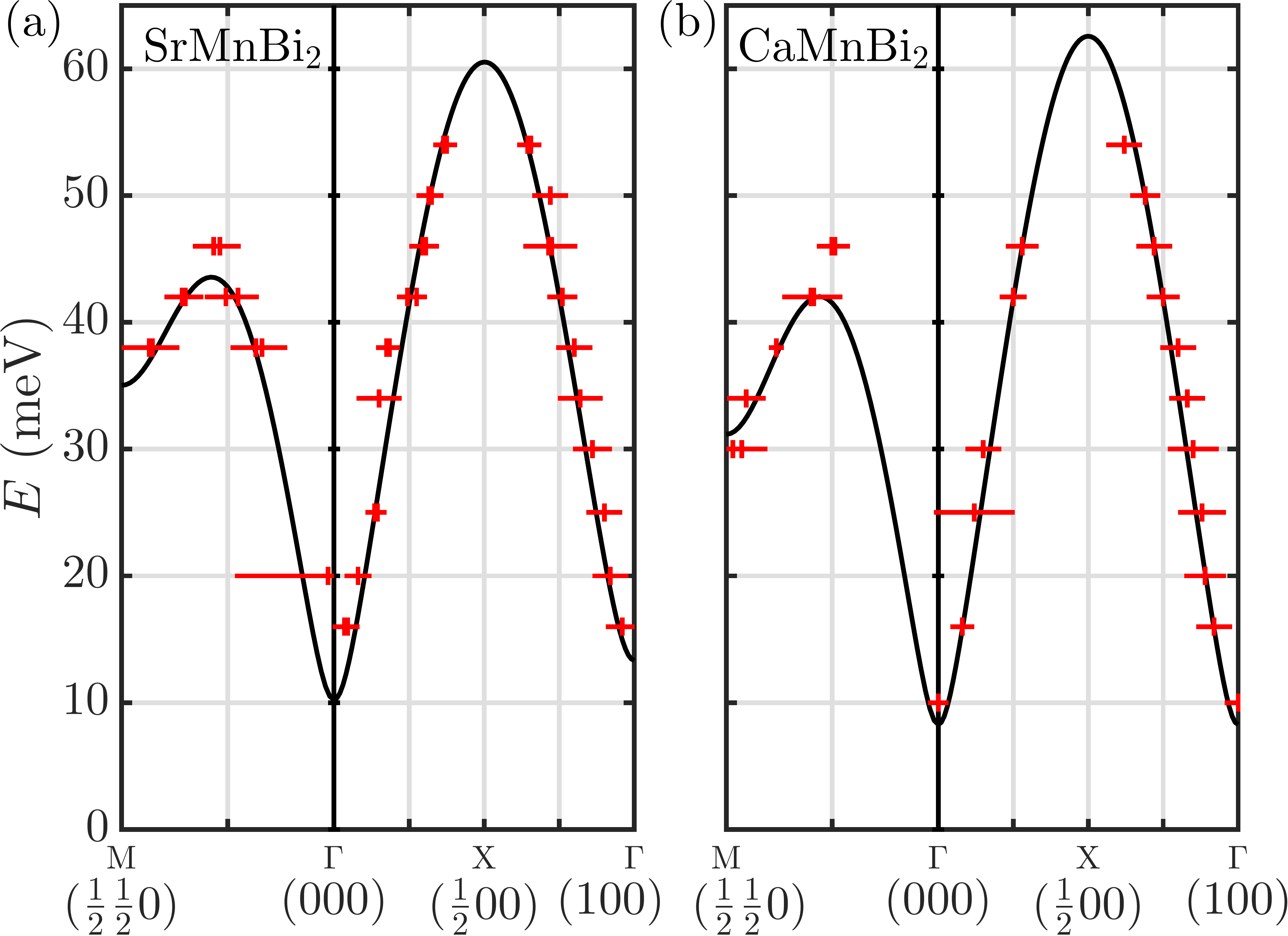}%
\caption{\label{fig5} (color online). Magnon dispersion along high symmetry directions in the $(HK0)$ plane, for (a) SrMnBi$_2$  and (b) CaMnBi$_2$. The black line indicates the best fit from the linear spin-wave model. Red markers represent the position (vertical bars) and full-width at half-maximum (horizontal lines) of gaussian fits to cuts through the raw data along the corresponding directions.}
\end{figure}

\begin{figure*}
\includegraphics[width=1.9\columnwidth,trim= 0pt 0pt 0pt 0pt, clip]{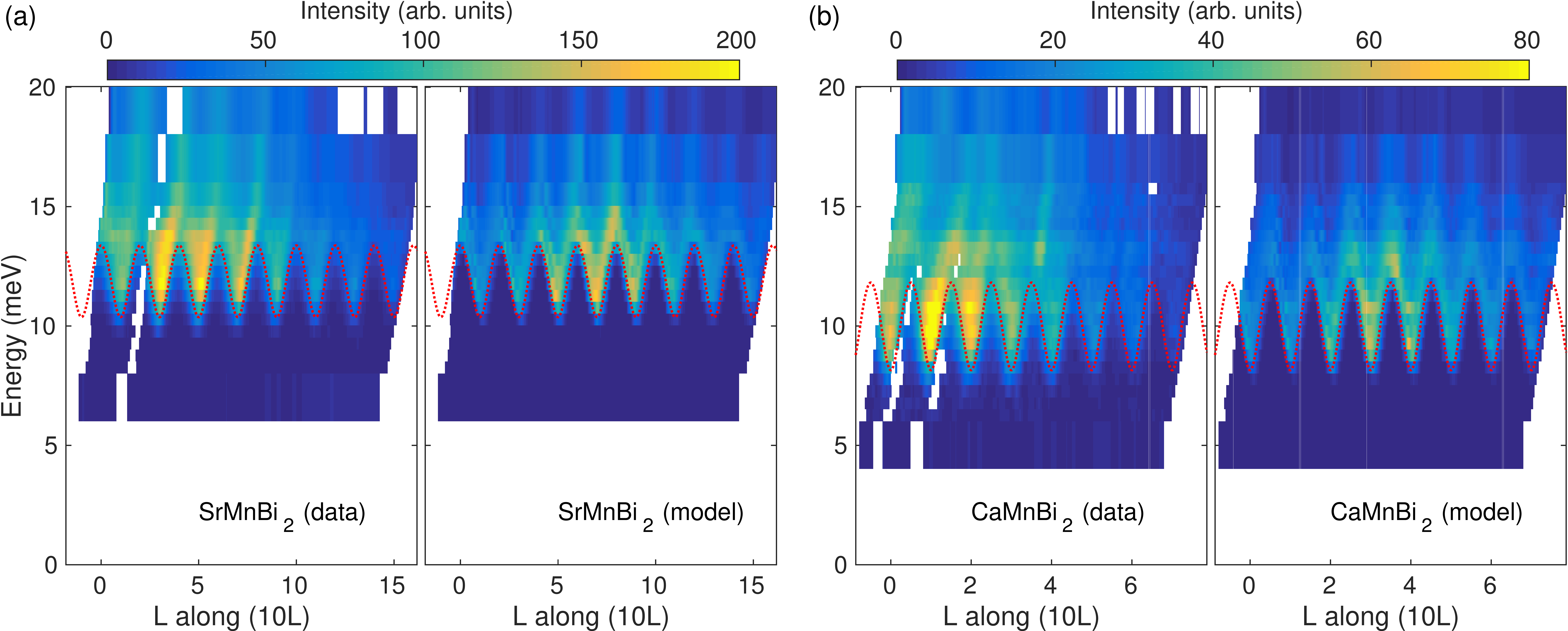}%
\caption{\label{fig6} (color online). Out-of-plane dispersion of the magnon spectra of (a) SrMnBi$_2$ and (b) CaMnBi$_2$, illustrated by slices along the $(10L)$ direction in reciprocal space (intensity averaged over the range $H=0.95$--1.05\,r.l.u.). The left-hand panels show interpolated plots of the data, and the right-hand panels give the corresponding best-fit spin-wave spectra convoluted with the instrumental resolution (for details, see text and Supplemental Material\cite{supplemental}). The superimposed red dashed line indicates the theoretical dispersion using best-fit parameters.}
\end{figure*}

\section{Results and analysis}

While cooling the samples in the $ac$ orientation, we tracked the intensities at selected positions in the $(H0L)$ plane of reciprocal space. Figures \ref{fig2}(a) and (b) show the resulting temperature dependences for SrMnBi$_2$ and CaMnBi$_2$, respectively. This includes the magnetic Bragg contribution at (101) and (103) (for Sr) and (100) (for Ca), as well as the diffuse magnetic scattering at another position along the $(10L)$ direction away from the Bragg condition. The data, here represented as the relative change in the square root of the intensity, demonstrate the order parameter characteristics of magnetic Bragg scattering at the antiferromagnetic transitions. We note that the decrease of the CaMnBi$_2$ $(100)$ magnetic scattering below 200$\,$K is not consistent with our previous powder neutron diffraction data,\cite{Guo2014} and could be due to a shift of the peak between two detector channels as the lattice contracts.

Above the ordering temperature, incipient in-plane magnetic correlations form diffuse rods of magnetic scattering along the $c^*$ direction of reciprocal space, as revealed at the $(10L)$ non-Bragg positions. When cooling towards $T_{\rm N}$, this diffuse scattering initially intensifies and then subsides when the weaker inter-layer correlations set in and neutron spectral weight is confined to the Bragg peaks. Fitting a power law to the thermal variation of the (101) (Sr) and (100) (Ca) peaks yields N\'{e}el temperatures of $T_{\rm N}^{\text{Sr}}=287(5)\,$K and $T_{\rm N}^{\text{Ca}}=264(2)\,$K. These values are consistent with previous single crystal bulk measurements of transport and ARPES samples \cite{Wang2011,Park2011,He2012,Wang2012}, but differ slightly from the values found in our earlier neutron powder diffraction study \cite{Guo2014}. This difference is likely due to small structural or compositional variations among the samples. The critical exponents $\beta^{\text{Sr}}=0.15(3)$ and $\beta^{\text{Ca}}=0.11(2)$ obtained from the power law fit are much smaller than the value $\beta=0.365$ of the three-dimensional Heisenberg model, indicating the reduced dimensionality of the magnetism in these systems. Due to the additional bismuth layers in the unit cells, the magnetism is more two-dimensional in {\it A}MnBi$_2$ than in the related (122) manganese arsenide BaMn$_2$As$_2$, $\beta=0.35(2)$ (Ref.~\onlinecite{Singh2009}). Instead, the inter-layer correlations compare well to the parent compounds of (122) iron-based superconductors, e.g.~$\beta=0.098(1)$ for SrFe$_2$As$_2$ (Ref.~\onlinecite{Tegel2008}) and $\beta=0.125$ for BaFe$_2$As$_2$ (Ref.~\onlinecite{Wilson2010}). A detailed description of the power law fit to this data is provided in the Supplemental Material.\cite{supplemental}

The measured neutron spectra are summarized in Fig.~\ref{fig3} (SrMnBi$_2$) and Fig.~\ref{fig4} (CaMnBi$_2$). A more quantitative presentation of the data is provided in figures S2--S5 of the Supplemental Material.\cite{supplemental} Due to the periodicity of the antiferromagnetic order, the magnetic zone centers are located at $(HKL)$ positions  with $(H+K)$ and $L$ both odd integers for SrMnBi$_2$, and at positions with $(H+K)$ odd and $L$ any integer for CaMnBi$_2$. For both compounds, the spectra reveal a well-defined magnon dispersion above spin gaps of approximately 10\,meV (Sr) and 8\,meV (Ca).  The magnons are highly dispersive parallel to the layers, but only weakly dispersive perpendicular to the layers. For both samples the magnon bandwidth is around 50\,meV for spin waves propagating in the $(HK0)$ plane and 3--4\,meV along $(10L)$. Figure~\ref{fig5} represents more quantitatively the magnon dispersion in the $(HK0)$ plane as obtained from gaussian fits to constant-energy cuts, and the left-hand panels of Figs.~\ref{fig6}(a) and (b) illustrate the out-of-plane dispersion by energy-wavevector slices of the data along the $(10L)$ direction.

To obtain quantitative information on the magnetic couplings we have compared the data with the linear spin-wave spectrum calculated from an effective spin Hamiltonian that includes a Heisenberg coupling term and an Ising-like single-ion anisotropy,
\begin{equation}
\mathcal{\hat{H}}=\sum_{\langle i,j \rangle}\,J_{ij}~\hat{\bf S}_i\cdot\hat{\bf S}_j\,-\,\sum_i\,D\,(\hat{S}_i^z)^2,
\end{equation}
where we include nearest-neighbor ($J_1$) and next-nearest-neighbour ($J_2$) exchange constants, an inter-layer exchange interaction $J_c$, and the anisotropy constant $D$. The exchange paths are shown in Fig.~\ref{fig1}. Using the Holstein--Primakoff transformation of two interacting Bose fields, corresponding to the two collinear antiferromagnetic sublattices, we obtain the dispersion relation
\begin{equation}\label{eq2}
E(\mathbf{q}) =  \sqrt{A(\mathbf{q})^2-B(\mathbf{q})^2}
\end{equation}
where $\bf q$ is the magnon wavevector,
\begin{align*}
A(\mathbf{q}) &= S\,\left[\, \mathcal{J}_{\mathrm{AF}}(0)-\mathcal{J}_{\mathrm{F}}(0)+\mathcal{J}_{\mathrm{F}}(\mathbf{q})+2\,D \,\right]\\
B(\mathbf{q}) &= S\, \,\mathcal{J}_{\mathrm{AF}}(\mathbf{q})
\end{align*}
and
\begin{equation}
\mathcal{J}(\mathbf{q})=\sum_n J_n\, {\rm e}^{2\pi{\rm i}\,\mathbf{q}\cdot\mathbf{r}_n}
\end{equation}
is the Fourier transform of the exchange interactions. The subscripts F and AF refer to summation over ferromagnetically- and antiferromagnetically-aligned spins, respectively.  The resulting differential scattering cross-section for single-magnon creation is
\begin{align}\label{eq1}
\frac{\mathrm{d}\sigma}{\mathrm{d}\Omega\mathrm{d}\omega}&=\frac{k_{\rm f}}{k_{\rm i}}\left(\frac{\gamma r_0}{2}\right)^{\hspace{-2pt} 2}\,S(\mathbf{q},\omega) \\
S(\mathbf{q},\omega) &=\,g^2 N S \,\frac{A(\mathbf{q})-B(\mathbf{q})}{E(\mathbf{q})}\, \{n(\omega)+1\}\, \delta\{\hbar \omega -E(\mathbf{q})\}
\label{eq2}
\end{align}
where $\hbar\omega$ is the neutron energy transfer, $k_{\rm f}$ and $k_{\bf i}$ are the outgoing and incoming neutron wave vectors, $\gamma=1.913$, $r_0$ is the classical electron radius, $g$ the Land\'{e} $g$-factor, $N$ the number of magnetic ions per sublattice, $S$ the spin quantum number, and $n(\omega)=({\rm e}^{\hbar\omega/k_{\mathrm{B}}T}-1)^{-1}$ the boson occupation number. Given the magnetic structures and exchange paths defined in Fig.~\ref{fig1}, the explicit Fourier exchange terms for the case of SrMnBi$_2$ are
\begin{align*}
\mathcal{J}_{\mathrm{AF}}^{\mathrm{Sr}}(\mathbf{q}) &= 2\,J_1\,\left[\cos\left(\pi\,H+\pi\,K\right)+\cos\left(\pi\,H-\pi\,K\right)\right] +\\
& ~~~~~+\, 2\,J_c\,\cos(\pi\,L)\\
\mathcal{J}_{\mathrm{F}}^{\mathrm{Sr}}(\mathbf{q}) &= 2\,J_2\,\left[\cos(2\pi\,H)+\cos(2\pi\,K)\right]
\end{align*}
and, in the case of CaMnBi$_2$,
\begin{align*}
\mathcal{J}_{\mathrm{AF}}^{\mathrm{Ca}}(\mathbf{q}) &= 2\,J_1\,\left[\cos\left(\pi\,H+\pi\,K\right)+\cos\left(\pi\,H-\pi\,K\right)\right] \\
\mathcal{J}_{\mathrm{F}}^{\mathrm{Ca}}(\mathbf{q}) &= 2\,J_2\,\left[\cos(2\pi\,H)+\cos(2\pi\,K)\right]+\, 2\,J_c\,\cos(2\pi\,L).
\end{align*}
This allows an analytical description of the spin gaps:
\begin{equation}\label{spingap}
\Delta^{\mathrm{Sr}} \, \approx \, \Delta^{\mathrm{Ca}} \, \approx \, 4\,\sqrt{SJ_1}\,\sqrt{SD},
\end{equation}
where we have applied the appropriate approximations for the present case ($J_1 \gg J_c$, $J_1\gg D$; for full expressions see the Supplemental Material\cite{supplemental}). Similarly, the band width $W$ of the dispersion along $(10L)$ is given by
\begin{equation}\label{bandwidth}
W^\mathrm{Sr} \, \approx \, W^{\mathrm{Ca}} \, \approx \, 4\,\sqrt{SJ_1}\,\left(\sqrt{SD+2\,|SJ_c|}-\sqrt{SD}\right).
\end{equation}

If $J_1$ is the dominant exchange, as is found to be the case here, then the maximum of the in-plane dispersion is $\sim 4SJ_1$. Given $J_1$, we see from Eqs.~(\ref{spingap}) and (\ref{bandwidth}) that in the relevant parameter regime the parameters $D$ and $J_c$ are determined by the size of the gap $\Delta^{\mathrm{Sr}/\mathrm{Ca}}$ and band width of the out-of-plane modulation $W^{\mathrm{Sr/Ca}}$, respectively. On the other hand, the balance between the parameters $J_1$ and $J_2$ determines details of the dispersion at higher energies in the $(HK0)$ plane. For example, a local minimum of the dispersion at the M point, $(\frac12,\frac12,0)$, as observed in both materials, will only occur for positive $J_2$, indicating a competition (frustration) between nearest- and next-nearest-neighbor exchange.

We find that the above model is able to reproduce very well all features in the data. For quantitative analysis we folded and averaged the raw constant-energy maps of reciprocal space into tiles of $2\times2$ r.l.u. With the data in this reduced form we could compare it to the model after convolution of the theoretical spectrum, Eqs.~(\ref{eq1})--(\ref{eq2}), with an energy- and wavevector-dependent broadening function to take into account the instrumental resolution.

A phenomenological gaussian broadening of the analytical dispersion proved insufficient to achieve a consistent global fit to the data, particularly for the low-energy part of the magnetic dispersion in the $ac$ plane. Instead, it was necessary to take into account the resolution of the triple-axis spectrometer, which was calculated with the RESTRAX ray-tracing algorithm\cite{Saroun1997,Saroun2002}. Our procedure to determine the parameters of the spin Hamiltonian $J_1$, $J_2$, $J_c$ and $D$ was carried out in three steps: First, a global fit of all data, using phenomenological gaussian broadening of the dispersion, produced rough estimates of all parameters. Using these as starting values and fixing the in-plane exchange interactions $J_1$ and $J_2$, we obtained precise bounds on the inter-layer exchange $J_c$ and anisotropy $D$ by fitting the resolution-convoluted spectrum for low energies (0--20\,meV) to an energy--wavevector slice with wavevector along $(10L)$, as illustrated in Fig.~\ref{fig6}. Finally, $J_1$ and $J_2$ were refined by fitting the in-plane ($ab$) dispersion at high energies (30--44\,meV) using gaussian broadening.

Figure \ref{fig5} provides a quantitative plot of the fits of the dispersions in the $ab$-plane. A more detailed description of the data processing, fitting and error estimation, and an explicit comparison of the data and best fits is provided in the Supplemental Material.\cite{supplemental}

\begin{table*}
\caption{\label{table1} Exchange parameters, magnetocrystalline anisotropy constants and spin gaps of SrMnBi$_2$ and CaMnBi$_2$ obtained from a fit of the linear spin-wave model, as described in the text. The parameters can be related to the nearest-neighbor ($d_\mathrm{NN}$) and interlayer ($d_c$) Mn--Mn atomic spacings, the ordered magnetic moment $\mu$, and the ordering temperature $T_\mathrm{N}$.\cite{Guo2014} The corresponding values for two related Mn pnictides are reproduced below.\cite{Calder2014,Johnston2011}}
\begin{ruledtabular}
\begin{tabular}{l|ccccc|cccc}
 & $SJ_1$ (meV) & $SJ_2$ (meV) & $SJ_c$ (meV) & $SD$ (meV) & $\Delta$ (meV) & $d_{\mathrm{NN}}$ (\AA) & $d_{\mathrm{c}}$ (\AA) & $\mu (\mu_\mathrm{B})$ & $T_\mathrm{N}$ (K) \\
\colrule
 SrMnBi$_2$ & 21.3(2) & 6.39(15)  & 0.11(2) & 0.31(2)  & 10.2(2) & 3.24 & 11.57 & 3.75(5) & 290.2(3) \\
 CaMnBi$_2$ & 23.4(6) & 7.9(5)  &-0.10(5) & 0.18(3)  & 8.3(8)  & 3.18 & 11.07 & 3.73(5) & 267.0(1.6) \\
\colrule
 BaMn$_2$Bi$_2$ [\onlinecite{Calder2014}] & 21.7(1.5) & 7.85(1.4) & 1.26(2) & 0.87(15)\footnote{The value of $SD$ for BaMn$_2$Bi$_2$ was misquoted in Ref.~\onlinecite{Calder2014}. In this table we give the correct value\cite{erratum}. } & 16.29(26) & 3.18 & 7.34 & 3.83(4) & 387.2(4)  \\
 BaMn$_2$As$_2$ [\onlinecite{Johnston2011}] & 33(3)  & 9.5(1.3)  & 3.0(6) & - & - &           2.95 & 6.73 & 3.88(4) & 625(1)  \\
\end{tabular}
\end{ruledtabular}
\end{table*}

\section{Discussion}

The exchange parameters for SrMnBi$_2$ and CaMnBi$_2$ obtained from the fits are summarized in Table~\ref{table1}. Apart from the opposite sign of the inter-layer exchange $J_c$, there are no significant differences between the parameters of SrMnBi$_2$ and CaMnBi$_2$. The absolute values of $J_1$ and $J_2$ are slightly larger in the case of CaMnBi$_2$, consistent with the smaller nearest-neighbor spacing ($d_\mathrm{NN}$). The magnitudes of $J_c$ for the two compounds, which are the same to within the error, are much smaller than $J_1$ and $J_2$, confirming the quasi-2D character of the magnetism in these materials.  Notably, these results are in good agreement with previous estimations based on first principles calculations of the electronic structure, which gave an average in-plane exchange of $SJ_{ab}\approx 30\,$meV and $|SJ_c|\approx 0.3\,$meV \cite{Guo2014}.

Regarding the magnetocrystalline anisotropy, we observe that $D$ is enhanced by a factor 1.8 in SrMnBi$_2$ compared with CaMnBi$_2$. According to the initial structure determinations at room temperature \cite{Cordier1977,Brechtel1980}, the local environment of the Mn ion is similar in both compounds: The MnBi$_4$ tetrahedra are elongated by $\approx$14$\,$\% along c and the ligand distances are $d^\mathrm{Ca}_\mathrm{Mn-Bi}=2.87(1)\,$\AA\, and $d^\mathrm{Sr}_\mathrm{Mn-Bi}=2.89(1)\,$\AA. The significant difference in anisotropy may therefore point to unknown structural distortions at 5$\,$K (at present, no full refinement of crystallographic parameters at low temperatures is available). The anisotropy in good agreement with the result of our earlier density functional prediction ($SD^\mathrm{Ca}_\mathrm{DFT}=0.3\,$meV \cite{Guo2014}), as was also the case with the exchange constants.

It is instructive to compare the present results to two available inelastic neutron studies of the related compounds BaMn$_2$Bi$_2$ \cite{Calder2014} and BaMn$_2$As$_2$ \cite{Johnston2011}. The corresponding parameters for these materials are quoted in Table~\ref{table1}. The pnictide-coordinated magnetic Mn$^{2+}$ layers in BaMn$_2$Bi$_2$ and BaMn$_2$As$_2$ (``122 materials'') are analogous to those in the 112 materials investigated in the present study. However, the $I4/mmm$ 122 compounds do not feature additional pnictide layers (which carry the Dirac bands in the present case). Hence, while the in-plane Mn--Mn spacing is very similar, the spacing of the magnetic layers in the 122 compounds is only 58--66\% of that in CaMnBi$_2$ and SrMnBi$_2$. Both BaMn$_2$Bi$_2$ and BaMn$_2$As$_2$ form antiferromagnetically stacked layers of N\'{e}el type order, in analogy to SrMnBi$_2$. As may be expected from these circumstances, we find that the in-plane exchange interactions in 122 compounds are similar or identical to those in 112 compounds. On the other hand, in the present 112 materials the inter-plane exchange is significantly reduced.
This is consistent with the much higher transition temperatures and the smaller separation of the Mn layers in the 122 materials compared with the 112 compounds.

We find no evidence that the additional Bi layers in 112 materials, which host the Dirac fermions, cause any qualitative changes in the magnon spectrum, such as anomalous broadening or dispersion. The instrument's simulated energy resolution provides an upper bound on the influence of such effects. The characteristics of the Bragg (0.5--$1.0\,$meV) and vanadium (1--$4\,$meV) widths of energy resolution are illustrated in the Supplemental Material.\cite{supplemental} By contrast, neutron inelastic measurements of many iron-based superconductors show obvious signatures of a strong hybridization of magnetic and itinerant states. A typical example is SrFe$_2$As$_2$ \cite{Ewings2011}, which shows a crossover into the regime of itinerant (\textit{Stoner}) spin fluctuations. This manifests itself as an increased dampening of spin fluctuations (i.e.~a broadening of the neutron spectrum) above a characteristic energy of approximately 80\,meV.

As in the 122 compounds, both $J_1$ and $J_2$ are positive (antiferromagnetic) in SrMnBi$_2$ and CaMnBi$_2$, resulting in frustration between nearest- and next-nearest-neighbor interactions. The theoretical phase diagram of the frustrated $J_1$--$J_2$ model on a square lattice has been investigated extensively in the context of iron-based superconductors.\cite{Mambrini2006,Richter2010,Wang2012a} There is special interest in this phase diagram  owing to a possible quantum critical point and spin liquid phase around $J_2/J_1\approx\frac12$. This regime separates two distinct ordered magnetic phases, with N\'{e}el type order for $J_2/J_1<\frac12$ and stripe antiferromagnetic order for $J_2/J_1>\frac12$. Both 112 and 122 Mn-based compounds exhibit dominant nearest-neighbor exchange, with $J_2/J_1\approx 0.3$. According to one study the exchange and anisotropy parameters for {\it A}MnBi$_2$ places these materials close to the phase boundary between N\'{e}el-ordered and frustrated paramagnetic phases.\cite{Wang2012a} The resulting quantum fluctuations could explain some of the observed reduction in ordered magnetic moment ($\simeq$3.7\,$\mu_\mathrm{B}$) compared to the ideal local-moment value of $5\,\mu_\mathrm{B}$ \cite{Johnston2011}.  By contrast, in parent compounds of iron-based superconductors such as BaFe$_2$As$_2$ and SrFe$_2$As$_2$, $J_1$ and $J_2$ are of similar magnitude, resulting in stripe-antiferromagnetic order.

\section{Conclusions}
In summary, we have performed a comprehensive triple-axis neutron scattering study of the anisotropic Dirac materials SrMnBi$_2$ and CaMnBi$_2$, with the aim of searching for possible influences of the unusual band topology at their Fermi surfaces on their magnetism. In particular, for CaMnBi$_2$ our previous findings had indicated that the Bi $6p_{x,y}$ bands may play a role in mediating the magnetic exchange between Mn layers.

In both compounds, we observed well-defined magnon spectra consistent with local-moment, semi-classical antiferromagnetism. Using linear spin-wave theory to describe the neutron spectra we have identified and quantified all relevant exchange and anisotropy parameters of a Heisenberg model for the two compounds. In both cases, all details of the dispersion are well reproduced by the model and there is no indication of anomalous broadening or dispersion to within experimental precision. The absolute values of the exchange parameters indicate no substantive differences between the compounds (aside from opposite interlayer coupling).

These results suggest that different routes have to be found to achieve an entanglement of magnetic order and non-trivial band topology. One very promising option is the substitution of magnetic rare earth ions on the \textit{A} site, providing a more direct interaction with the relevant Bi layers. In particular, a strong response of the transport properties to rare earth magnetic order has recently been observed in EuMnBi$_2$ \cite{May2014}, along with the trademark signatures of Dirac transport \cite{Masuda2016}. Furthermore, recent high-resolution ARPES results and first principles calculations identify YbMnBi$_2$ as a type-2 Weyl semimetal with canted antiferromagnetic order \cite{Borisenko2016}. The latter study further suggests that this state would be tuned to a Dirac metal by spin alignment. Naturally, it would be of great interest to perform analogous inelastic neutron studies of the magnetic ground states in those materials.

Note added in proof: After submission of this manuscript, a Raman spectroscopic study of SrMnBi$_2$ and CaMnBi$_2$ was reported by Zhang \textit{et al.} \cite{ZhangLiuYiEtAl2016}. Raman spectroscopy probes the spin dynamics through a small number of characteristic frequencies which are associated with van-Hove singularities in the two-magnon density of states. The authors of Ref.~\onlinecite{ZhangLiuYiEtAl2016} interpret their data using a similar spin Hamiltonian as in the present study but without the magnetocrystalline anisotropy term ($D$ in our study). Their analysis yields values for the spin exchange parameters $J_1$ and $J_2$ that are similar to our results, but produces anomalously large values of the inter-layer exchange $J_c$ for both materials (one order of magnitude larger than in our study or in other related materials). The authors of Ref.~\onlinecite{ZhangLiuYiEtAl2016} suggest that this enhanced coupling is caused by the Bi Dirac bands.  We would like to draw attention to the fact that the parameters $J_1$, $J_c$ and $D$ are strongly correlated in modelling key features of the magnon dispersion (see Eqs.~\ref{spingap} and \ref{bandwidth}), so the neglect of $D$ in the Raman analysis could significantly affect the obtained values of $J_c$.  We note that the Raman value of $J_c$ would imply an inter-layer dispersion of the one-magnon spectrum at a factor of 11 (Sr) or 7 (Ca) larger than that found here directly by neutron spectroscopy (Fig.~\ref{fig6}, Eq. \ref{bandwidth}).

\begin{acknowledgments}
We would like to thank Dr. Paul Steffens (ILL) for providing the FLATCONE data treatment suite, and Dr Stuart Calder for helpful clarifications on the results reported in Ref.~\onlinecite{Calder2014}. This work was supported by the U.K. Engineering and Physical Sciences Research Council (grant no. EP/J017124/1), the Chinese National Key Research and Development Program (2016YFA0300604) and the Strategic Priority Research Program (B) of the Chinese Academy of Sciences (Grant No. XDB07020100). MCR is grateful to the Oxford University Clarendon Fund for provision of a scholarship.
\end{acknowledgments}

\bibliography{AMnBi2Bib}

\cleardoublepage
\onecolumngrid
\appendix

\begin{center}

\Large
Supplemental Material:

\vspace{1cm}

\large
{\bf Spin dynamics in the antiferromagnetic phases of the\\ Dirac metals $A$MnBi$_2$ ($A=$ Sr, Ca)}

\vspace{1cm}

\normalsize

M. C. Rahn,$^1$ A. J. Princep,$^1$ A. Piovano,$^2$ J. Kulda,$^2$ Y. F. Guo,$^{3,4}$ Y. G. Shi,$^5$ and A. T. Boothroyd$^1$

\vspace{0.2cm}
\small

$^1${\it Clarendon Laboratory, Department of Physics, University of Oxford, Oxford, OX1 3PU, United Kingdom}\\[1pt]
$^2${\it Institut Laue-Langevin, 6 rue Jules Horowitz, 38042 Grenoble Cedex 9, France}\\[1pt]
$^3${\it School of Physical Science and Technology, ShanghaiTech University, Shanghai 201210, China}\\[1pt]
$^4${\it CAS Center for Excellence in Superconducting Electronics (CENSE), Shanghai 200050, China}\\[1pt]
$^5${\it Beijing National Laboratory for Condensed Matter Physics, Institute of Physics, Chinese Academy of Sciences,\\ Beijing 100190, China}\\

\end{center}
\vspace{1cm}

\twocolumngrid

\begin{center}
{\bf 1. Power law fit of magnetic Bragg peaks}
\end{center}

As described in the manuscript, the thermal variation of the intensities of the (101) (SrMnBi$_2$) and (100) (CaMnBi$_2$) magnetic peaks were fitted with a power law. To reproduce the incipient fluctuations above $T_\mathrm{N}$, the power law fitting function was convoluted by a Lorentzian distribution of the ordering temperature:
\begin{equation*}
I\,\propto\,A\,(T_\mathrm{N}-T)^{2\beta}\,\,\ast\,\,\frac{1}{\gamma^2+(T-T_\mathrm{N})^2}\,.
\end{equation*}
The five (unconstrained) fitting parameters where then (1) a constant background, (2) an overall scale factor $A$, (3) the critical exponent $\beta$, (4) the N\'eel temperature $T_\mathrm{N}$ and (5) the Lorentzian full width at half maximum $\gamma$ (see Fig. S1).
While $T_\mathrm{N}$ always converges to the results quoted in the main text, the critical exponents $\beta$ weakly depend on the selected fitting range. In the main text we therefore quote the mean fit results and standard deviations of separate fits to data ranges varying between ($T_\mathrm{N}-65\,$K) to $313\,$K and ($T_\mathrm{N}-5\,$K) to $313\,$K.

\begin{figure}[H]
\includegraphics[width=0.90\columnwidth,trim= 0pt 0pt 0pt 0pt, clip]{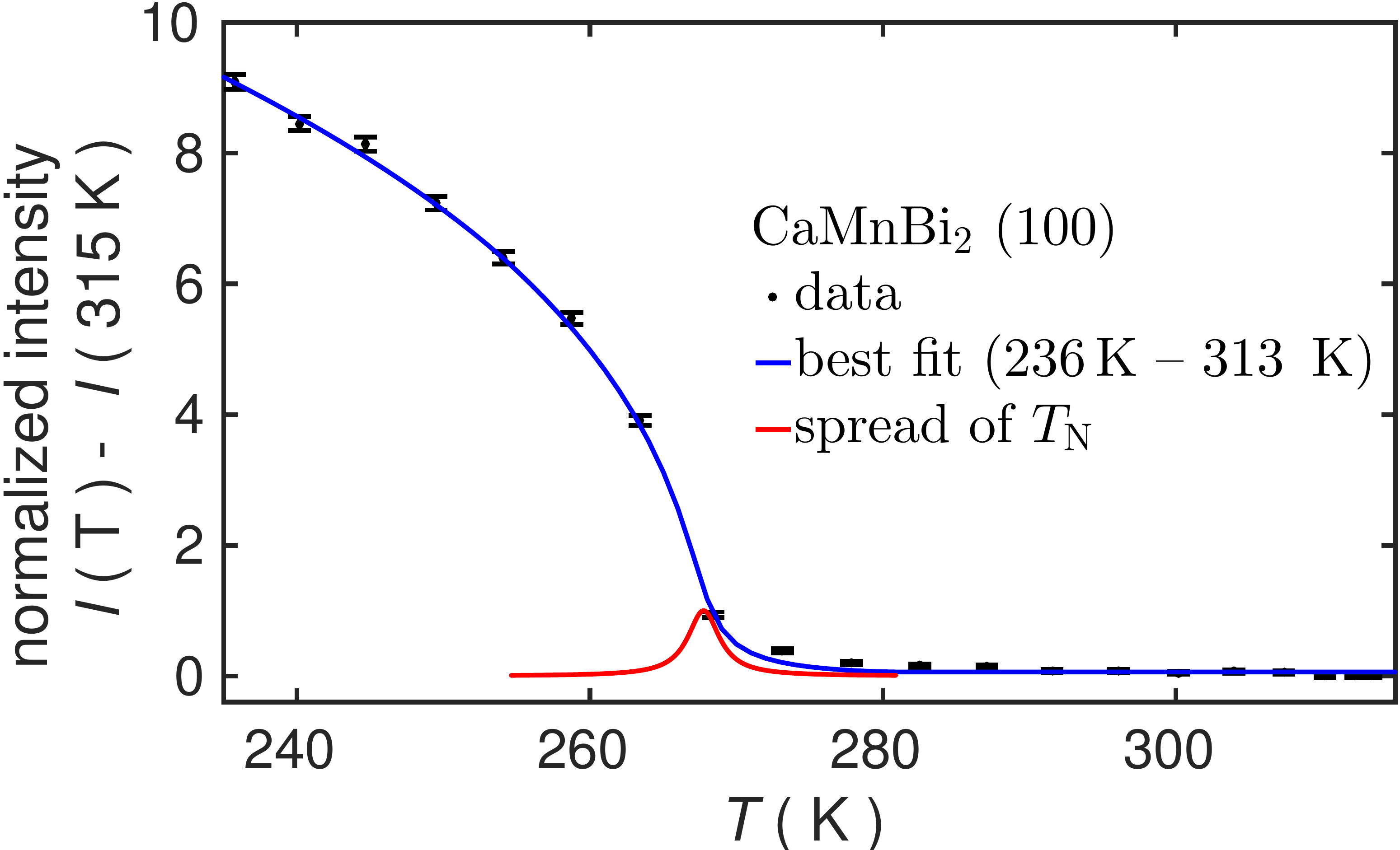}%
\justify{\label{sfig-1} Fig.~S1 (color online). Lorentzian-convoluted power law fit to the relative thermal variation of the scattering amplitude at the magnetic Bragg reflection (100) of CaMnBi$_2$.}
\end{figure}

\vspace{5pt}

\begin{center}
{\bf 2. Magnetic susceptibility}
\end{center}

\begin{figure}[b]
\includegraphics[width=0.95\columnwidth,trim= 0pt 0pt 0pt 0pt, clip]{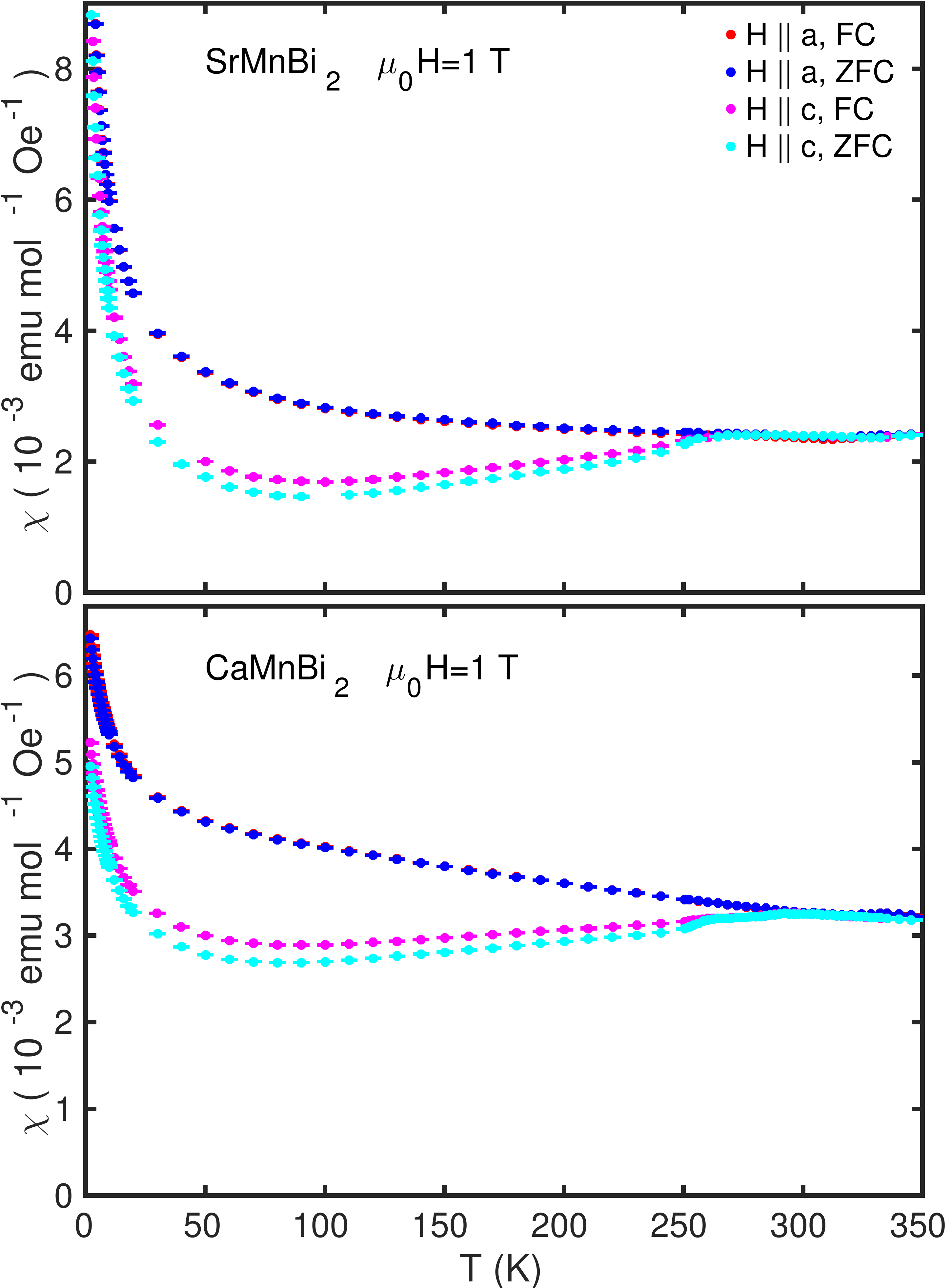}%
\justify{\label{sfig0} Fig.~S2 (color online). Magnetic susceptibility of SrMnBi$_2$ (top) and CaMnBi$_2$ (bottom), measured on the batch of single crystals probed by neutron spectroscopy.}
\end{figure}
To check for consistency with our previous study \cite{Guo2014}, we measured the thermal variation of the magnetic properties of the batch of samples probed by neutron inelastic scattering (see Fig.~S2). Single crystals were first aligned on an x-ray diffractometer. The magnetic susceptibility was then measured on a Magnetic Properties Measurement System (Quantum Design) with a magnetic field of flux density 1\,Tesla applied either in-plane or out-of-plane. The key characteristics are qualitatively consistent with our earlier results \citep{Guo2014}. The present samples have a larger Curie contribution which may be attributed to paramagnetic impurities induced by sample decay. Since the magnon dispersion is a coherent response of the main crystal phase, such impurities are not of relevance to the present study, apart from a small contribution to the diffuse background scattering.


\begin{center}
{\bf 3. Analytical expressions for features in the magnon dispersion}
\end{center}

The linear spin-wave model yields the following analytical expressions for the spin gaps:
\begin{align*}
\Delta^{\mathrm{Sr}} &=\, S\,\left[ (4\,J_1 + 2\,J_c + 2\, D)^2-(4\,J_1 + 2\, J_c)^2\right]^{\frac12}\\[5pt]
					& \approx 4\,\sqrt{SJ_1}\,\sqrt{SD}\\[10pt]
\Delta^{\mathrm{Ca}} &=\, S\,\left[ (4\,J_1 + 2\, D)^2-(4\,J_1)^2\right]^{\frac12} \\[5pt]
					& \approx 4\,\sqrt{SJ_1}\,\sqrt{SD}\,,
\end{align*}
where we have applied the appropriate approximations for the present case ($J_1 \gg J_c$; $J_1\gg D$). Similarly, for the bandwidth $W$ of the dispersion along $(10L)$ we obtain
\begin{align*}
W^\mathrm{Sr} &=\, S\, \left[ (4J_1+2J_c+2D)^2-(4J_1-2J_c)^2 \right]^\frac12  - \\
 &~~~-S\,\left[ (4J_1+2J_c+2D)^2-(4J_1+2J_c)^2 \right]^\frac12  \\[5pt]
 & \approx 4\,\sqrt{SJ_1}\,\left(\sqrt{SD+2\,SJ_c}-\sqrt{SD}\right)\\[10pt]
W^\mathrm{Ca} &=\, S\, \left[ (4J_1-4J_c+2D)^2-(4J_1)^2 \right]^\frac12  - \\
 &~~~-S\,\left[ (4J_1+2D)^2-(4J_1)^2 \right]^\frac12 \\[5pt]
 & \approx 4\,\sqrt{SJ_1}\,\left(\sqrt{SD-2\,SJ_c}-\sqrt{SD}\right)\,.\\
\end{align*}


\begin{center}
{\bf 4. Analysis of neutron spectra}
\end{center}

For both materials, and for each in two crystal orientations (momentum transfers in the $(HK0)$ and $(H0L)$ planes of reciprocal space), rocking scans were recorded at a number of energy transfers up to $60$\,meV. The scattering was recorded in the FlatCone analyser--detector assembly \cite{Kempa2006} producing a set of constant-energy maps. In each rocking scan the wide FlatCone detector bank is mapped onto an arc in reciprocal space intercepting several Brillouin zones. The spectral weight in these maps shows contributions from Bragg diffraction (at $E=0\,$meV), acoustic phonons (below $E\approx 10\,$meV) and magnons (above $E=8$--$10$\,meV). In addition, for $E\neq 0\,$meV, the data contain accidental Bragg reflections (i.e. the strong elastic signal passes through the analyzer by parasitic scattering), as well as powder rings corresponding to Bragg diffraction by the sample environment. The magnon spectral weight can clearly be distinguished from the phonon- and accidental Bragg scattering in every dataset from the form of the scattering and from the reduction in the magnetic signal with $|{\bf q}|$.

To allow a direct comparison with the model, each raw dataset was corrected in several steps. First, all intensity other than a diffuse background and aluminium powder rings were masked. This masked data was then radially averaged (along the rocking angle). Next, both the raw dataset and the radially averaged powder/background dataset were divided by the magnetic form factor of Mn$^{2+}$. After subtraction, the trajectories of intense spurious (parasitic) Bragg scattering were deleted from the data manually. The resulting corrected constant-energy maps were then interpolated to a regular grid and divided into equivalent tiles of $2 \times 2$ reciprocal lattice units (r.l.u.). Finally, the arithmetic mean of these cells was calculated and all relevant symmetries (mirror planes and 4- or 2-fold rotation of the $(HK0)$ and $(H0L)$ reciprocal lattice planes, respectively) were applied in order to distil the full statistical significance of the data. The resulting constant-energy intensity maps for both compounds and both sample orientations are shown in Figs.~S4--S7.


\begin{center}
{\bf 5. Fitting procedure}
\end{center}

 The constant-energy maps thus obtained are in a convenient form for fitting to the spin-wave model. However, in particular for the low energy part of the magnetic dispersion, a phenomenological gaussian broadening of the analytical dispersion (along the reciprocal space and energy dimensions) proved insufficient. Instead, it was necessary to take into account the resolution of the triple-axis spectrometer. This was done by numerically simulating four-dimensional resolution matrices using the RESTRAX ray tracing algorithm \cite{Saroun1997,Saroun2002}. This algorithm takes into account a full physical model of all relevant components (including collimators, monochromator, analyzer, detectors) of the IN8 instrument. To calculate the resolution matrix for a particular instrument configuration, the program was set to trace 5000 random neutron events. The resulting cloud of $(H,K,L,E)$ positions was fitted to a four-dimensional ellipsoid. 
 
The energy resolution of the triple-axis spectrometer depends only weakly on the momentum transfer, but increases with energy transfer. While the energy Bragg-width of the resolution ellipsoid remains smaller than $1\,$meV (standard deviation) up to $\Delta E=60\,$meV, the vanadium width increases up to $\approx4\,$meV in this range. Figure S3 illustrates these characteristics, which represent the upper bounds within which we can exclude anomalies in the dispersion.

\begin{figure}[H]
\includegraphics[width=0.90\columnwidth,trim= 0pt 0pt 0pt 0pt, clip]{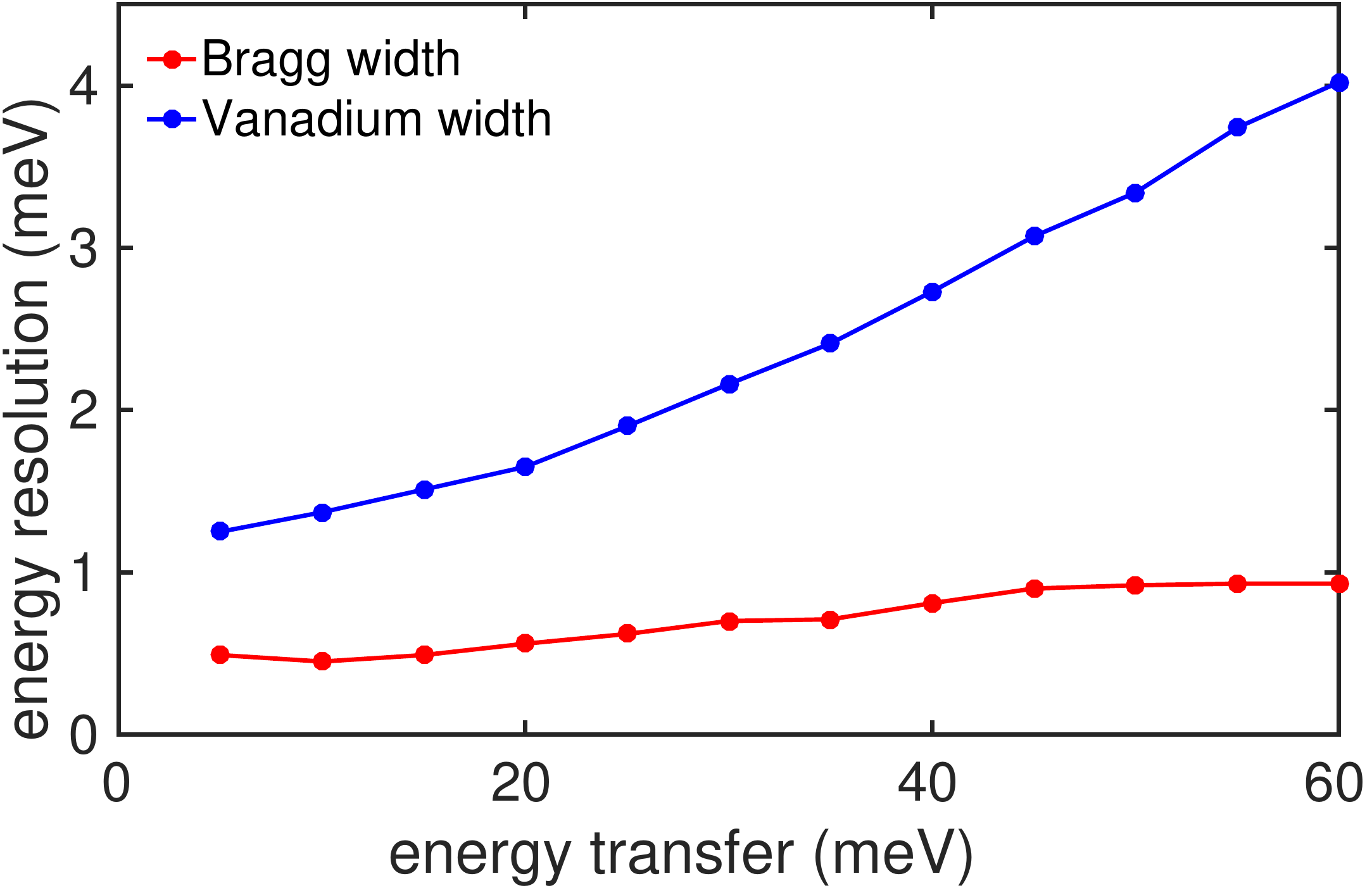}%
\justify{\label{sfig-1} Fig.~S3 (color online). Simulated energy resolution of the IN8 spectrometer (RESTRAX algorithm) for the experimental conditions of the present study.}
\end{figure}

 In order to simulate the effect of the resolution on the distribution of magnetic intensity, the resolution ellipsoids were calculated for every pixel of a full dataset. For a particular set of model parameters, the contribution to the intensity at a pixel was calculated by performing a four-dimensional convolution of the ellipsoid with the analytical dispersion. Finally, the so-obtained resolution-corrected intensity corresponding to the full experimental dataset was folded into $2 \times 2$\,r.l.u. tiles as described for the raw data above.

While this procedure produces a satisfactory simulation of the data, it is unfortunately computationally too intensive to include in an efficient global fitting routine. The best-fit parameters were therefore determined in three stages: First, a global fit (with phenomenological broadening) of all available datasets (all energies, both orientations) was used to estimate rough starting values for all parameters. Secondly, for an appropriate grid of $J_c$ and $D$ values, resolution-convoluted constant-energy maps corresponding to the low energy ($E\leq 20\,$meV) data sets were calculated. Momentum transfer slices along the $(10L)$ direction were then calculated for both data and model. The $\chi^2$ maps resulting from this comparison are shown in Figs.~S8(a) and (c) for SrMnBi$_2$ and CaMnBi$_2$, respectively, with the contour of one standard deviation indicated by a red line. The best fit $(10L)$ slices thus determined are illustrated in Fig.~6 of the main article.

Finally, with $J_c$ and $D$ fixed, global fits of the high energy dispersion ($30\leq E\leq 45\,$\,meV, both $HK0$ and $H0L$ orientations) were performed for an appropriate grid of $J_1$ and $J_2$ values, using a phenomenological broadening of the dispersion. The resulting $\chi^2$ maps are shown in Fig.~S8(b) and (d) for SrMnBi$_2$ and CaMnBi$_2$, respectively. The $J_1$ and $J_2$ parameters are seen to be strongly correlated in both cases, with a minimum in $\chi^2$ extending along a straight line.  However, although $\chi^2$ does not vary significantly along this line, there are weak features in the high energy $(HK0)$ maps which do depend sensitively on $(J_1, J_2)$ along this line. These features enabled us to establish empirically the errors on $J_1$ and $J_2$ quoted in the main article and indicated by dashed lines in Fig.~S8.

Figures~S4--S7 give a comparison of all the data with the corresponding simulations from the best-fit model.

\bibliography{AMnBi2Bib}

\begin{thebibliography}{32}%
\makeatletter
\providecommand \@ifxundefined [1]{%
 \@ifx{#1\undefined}
}%
\providecommand \@ifnum [1]{%
 \ifnum #1\expandafter \@firstoftwo
 \else \expandafter \@secondoftwo
 \fi
}%
\providecommand \@ifx [1]{%
 \ifx #1\expandafter \@firstoftwo
 \else \expandafter \@secondoftwo
 \fi
}%
\providecommand \natexlab [1]{#1}%
\providecommand \enquote  [1]{``#1''}%
\providecommand \bibnamefont  [1]{#1}%
\providecommand \bibfnamefont [1]{#1}%
\providecommand \citenamefont [1]{#1}%
\providecommand \href@noop [0]{\@secondoftwo}%
\providecommand \href [0]{\begingroup \@sanitize@url \@href}%
\providecommand \@href[1]{\@@startlink{#1}\@@href}%
\providecommand \@@href[1]{\endgroup#1\@@endlink}%
\providecommand \@sanitize@url [0]{\catcode `\\12\catcode `\$12\catcode
  `\&12\catcode `\#12\catcode `\^12\catcode `\_12\catcode `\%12\relax}%
\providecommand \@@startlink[1]{}%
\providecommand \@@endlink[0]{}%
\providecommand \url  [0]{\begingroup\@sanitize@url \@url }%
\providecommand \@url [1]{\endgroup\@href {#1}{\urlprefix }}%
\providecommand \urlprefix  [0]{URL }%
\providecommand \Eprint [0]{\href }%
\providecommand \doibase [0]{http://dx.doi.org/}%
\providecommand \selectlanguage [0]{\@gobble}%
\providecommand \bibinfo  [0]{\@secondoftwo}%
\providecommand \bibfield  [0]{\@secondoftwo}%
\providecommand \translation [1]{[#1]}%
\providecommand \BibitemOpen [0]{}%
\providecommand \bibitemStop [0]{}%
\providecommand \bibitemNoStop [0]{.\EOS\space}%
\providecommand \EOS [0]{\spacefactor3000\relax}%
\providecommand \BibitemShut  [1]{\csname bibitem#1\endcsname}%
\let\auto@bib@innerbib\@empty
\bibitem [{\citenamefont {Cordier}\ and\ \citenamefont
  {Sch{\"a}fer}(1977)}]{Cordier1977}%
  \BibitemOpen
  \bibfield  {author} {\bibinfo {author} {\bibfnamefont {G.}~\bibnamefont
  {Cordier}}\ and\ \bibinfo {author} {\bibfnamefont {H.}~\bibnamefont
  {Sch{\"a}fer}},\ }\href@noop {} {\bibfield  {journal} {\bibinfo  {journal}
  {Z. Naturforsch.}\ }\textbf {\bibinfo {volume} {32}},\ \bibinfo {pages} {383}
  (\bibinfo {year} {1977})}\BibitemShut {NoStop}%
\bibitem [{\citenamefont {Brechtel}\ \emph {et~al.}(1980)\citenamefont
  {Brechtel}, \citenamefont {Cordier},\ and\ \citenamefont
  {Sch{\"a}fer}}]{Brechtel1980}%
  \BibitemOpen
  \bibfield  {author} {\bibinfo {author} {\bibfnamefont {E.}~\bibnamefont
  {Brechtel}}, \bibinfo {author} {\bibfnamefont {G.}~\bibnamefont {Cordier}}, \
  and\ \bibinfo {author} {\bibfnamefont {H.}~\bibnamefont {Sch{\"a}fer}},\
  }\href
  {//www.degruyter.com/view/j/znb.1980.35.issue-1/znb-1980-0102/znb-1980-0102.xml}
  {\bibfield  {journal} {\bibinfo  {journal} {Z. Naturforsch.}\ }\textbf
  {\bibinfo {volume} {35}},\ \bibinfo {pages} {1} (\bibinfo {year}
  {1980})}\BibitemShut {NoStop}%
\bibitem [{\citenamefont {Shim}\ \emph {et~al.}(2009)\citenamefont {Shim},
  \citenamefont {Haule},\ and\ \citenamefont {Kotliar}}]{Shim2009}%
  \BibitemOpen
  \bibfield  {author} {\bibinfo {author} {\bibfnamefont {J.~H.}\ \bibnamefont
  {Shim}}, \bibinfo {author} {\bibfnamefont {K.}~\bibnamefont {Haule}}, \ and\
  \bibinfo {author} {\bibfnamefont {G.}~\bibnamefont {Kotliar}},\ }\href
  {\doibase 10.1103/PhysRevB.79.060501} {\bibfield  {journal} {\bibinfo
  {journal} {Phys. Rev. B}\ }\textbf {\bibinfo {volume} {79}},\ \bibinfo
  {pages} {060501} (\bibinfo {year} {2009})}\BibitemShut {NoStop}%
\bibitem [{\citenamefont {Wang}\ \emph {et~al.}(2011)\citenamefont {Wang},
  \citenamefont {Zhao}, \citenamefont {Yin}, \citenamefont {Kotliar},
  \citenamefont {Kim}, \citenamefont {Aronson},\ and\ \citenamefont
  {Morosan}}]{Wang2011}%
  \BibitemOpen
  \bibfield  {author} {\bibinfo {author} {\bibfnamefont {J.~K.}\ \bibnamefont
  {Wang}}, \bibinfo {author} {\bibfnamefont {L.~L.}\ \bibnamefont {Zhao}},
  \bibinfo {author} {\bibfnamefont {Q.}~\bibnamefont {Yin}}, \bibinfo {author}
  {\bibfnamefont {G.}~\bibnamefont {Kotliar}}, \bibinfo {author} {\bibfnamefont
  {M.~S.}\ \bibnamefont {Kim}}, \bibinfo {author} {\bibfnamefont {M.~C.}\
  \bibnamefont {Aronson}}, \ and\ \bibinfo {author} {\bibfnamefont
  {E.}~\bibnamefont {Morosan}},\ }\href {\doibase 10.1103/PhysRevB.84.064428}
  {\bibfield  {journal} {\bibinfo  {journal} {Phys. Rev. B}\ }\textbf {\bibinfo
  {volume} {84}},\ \bibinfo {pages} {064428} (\bibinfo {year}
  {2011})}\BibitemShut {NoStop}%
\bibitem [{\citenamefont {Ran}\ \emph {et~al.}(2009)\citenamefont {Ran},
  \citenamefont {Wang}, \citenamefont {Zhai}, \citenamefont {Vishwanath},\ and\
  \citenamefont {Lee}}]{Ran2009}%
  \BibitemOpen
  \bibfield  {author} {\bibinfo {author} {\bibfnamefont {Y.}~\bibnamefont
  {Ran}}, \bibinfo {author} {\bibfnamefont {F.}~\bibnamefont {Wang}}, \bibinfo
  {author} {\bibfnamefont {H.}~\bibnamefont {Zhai}}, \bibinfo {author}
  {\bibfnamefont {A.}~\bibnamefont {Vishwanath}}, \ and\ \bibinfo {author}
  {\bibfnamefont {D.-H.}\ \bibnamefont {Lee}},\ }\href {\doibase
  10.1103/PhysRevB.79.014505} {\bibfield  {journal} {\bibinfo  {journal} {Phys.
  Rev. B}\ }\textbf {\bibinfo {volume} {79}},\ \bibinfo {pages} {014505}
  (\bibinfo {year} {2009})}\BibitemShut {NoStop}%
\bibitem [{\citenamefont {Richard}\ \emph {et~al.}(2010)\citenamefont
  {Richard}, \citenamefont {Nakayama}, \citenamefont {Sato}, \citenamefont
  {Neupane}, \citenamefont {Xu}, \citenamefont {Bowen}, \citenamefont {Chen},
  \citenamefont {Luo}, \citenamefont {Wang}, \citenamefont {Dai}, \citenamefont
  {Fang}, \citenamefont {Ding},\ and\ \citenamefont {Takahashi}}]{Richard2010}%
  \BibitemOpen
  \bibfield  {author} {\bibinfo {author} {\bibfnamefont {P.}~\bibnamefont
  {Richard}}, \bibinfo {author} {\bibfnamefont {K.}~\bibnamefont {Nakayama}},
  \bibinfo {author} {\bibfnamefont {T.}~\bibnamefont {Sato}}, \bibinfo {author}
  {\bibfnamefont {M.}~\bibnamefont {Neupane}}, \bibinfo {author} {\bibfnamefont
  {Y.-M.}\ \bibnamefont {Xu}}, \bibinfo {author} {\bibfnamefont {J.~H.}\
  \bibnamefont {Bowen}}, \bibinfo {author} {\bibfnamefont {G.~F.}\ \bibnamefont
  {Chen}}, \bibinfo {author} {\bibfnamefont {J.~L.}\ \bibnamefont {Luo}},
  \bibinfo {author} {\bibfnamefont {N.~L.}\ \bibnamefont {Wang}}, \bibinfo
  {author} {\bibfnamefont {X.}~\bibnamefont {Dai}}, \bibinfo {author}
  {\bibfnamefont {Z.}~\bibnamefont {Fang}}, \bibinfo {author} {\bibfnamefont
  {H.}~\bibnamefont {Ding}}, \ and\ \bibinfo {author} {\bibfnamefont
  {T.}~\bibnamefont {Takahashi}},\ }\href {\doibase
  10.1103/PhysRevLett.104.137001} {\bibfield  {journal} {\bibinfo  {journal}
  {Phys. Rev. Lett.}\ }\textbf {\bibinfo {volume} {104}},\ \bibinfo {pages}
  {137001} (\bibinfo {year} {2010})}\BibitemShut {NoStop}%
\bibitem [{\citenamefont {Morinari}\ \emph {et~al.}(2010)\citenamefont
  {Morinari}, \citenamefont {Kaneshita},\ and\ \citenamefont
  {Tohyama}}]{Morinari2010}%
  \BibitemOpen
  \bibfield  {author} {\bibinfo {author} {\bibfnamefont {T.}~\bibnamefont
  {Morinari}}, \bibinfo {author} {\bibfnamefont {E.}~\bibnamefont {Kaneshita}},
  \ and\ \bibinfo {author} {\bibfnamefont {T.}~\bibnamefont {Tohyama}},\ }\href
  {\doibase 10.1103/PhysRevLett.105.037203} {\bibfield  {journal} {\bibinfo
  {journal} {Phys. Rev. Lett.}\ }\textbf {\bibinfo {volume} {105}},\ \bibinfo
  {pages} {037203} (\bibinfo {year} {2010})}\BibitemShut {NoStop}%
\bibitem [{\citenamefont {Vafek}\ and\ \citenamefont
  {Vishwanath}(2014)}]{Vafek2014}%
  \BibitemOpen
  \bibfield  {author} {\bibinfo {author} {\bibfnamefont {O.}~\bibnamefont
  {Vafek}}\ and\ \bibinfo {author} {\bibfnamefont {A.}~\bibnamefont
  {Vishwanath}},\ }\href@noop {} {\bibfield  {journal} {\bibinfo  {journal}
  {Annu. Rev. Cond. Mat. Phys.}\ }\textbf {\bibinfo {volume} {5}},\ \bibinfo
  {pages} {83} (\bibinfo {year} {2014})}\BibitemShut {NoStop}%
\bibitem [{\citenamefont {Park}\ \emph {et~al.}(2011)\citenamefont {Park},
  \citenamefont {Lee}, \citenamefont {Wolff-Fabris}, \citenamefont {Koh},
  \citenamefont {Eom}, \citenamefont {Kim}, \citenamefont {Farhan},
  \citenamefont {Jo}, \citenamefont {Kim}, \citenamefont {Shim},\ and\
  \citenamefont {Kim}}]{Park2011}%
  \BibitemOpen
  \bibfield  {author} {\bibinfo {author} {\bibfnamefont {J.}~\bibnamefont
  {Park}}, \bibinfo {author} {\bibfnamefont {G.}~\bibnamefont {Lee}}, \bibinfo
  {author} {\bibfnamefont {F.}~\bibnamefont {Wolff-Fabris}}, \bibinfo {author}
  {\bibfnamefont {Y.~Y.}\ \bibnamefont {Koh}}, \bibinfo {author} {\bibfnamefont
  {M.~J.}\ \bibnamefont {Eom}}, \bibinfo {author} {\bibfnamefont {Y.~K.}\
  \bibnamefont {Kim}}, \bibinfo {author} {\bibfnamefont {M.~A.}\ \bibnamefont
  {Farhan}}, \bibinfo {author} {\bibfnamefont {Y.~J.}\ \bibnamefont {Jo}},
  \bibinfo {author} {\bibfnamefont {C.}~\bibnamefont {Kim}}, \bibinfo {author}
  {\bibfnamefont {J.~H.}\ \bibnamefont {Shim}}, \ and\ \bibinfo {author}
  {\bibfnamefont {J.~S.}\ \bibnamefont {Kim}},\ }\href {\doibase
  10.1103/PhysRevLett.107.126402} {\bibfield  {journal} {\bibinfo  {journal}
  {Phys. Rev. Lett.}\ }\textbf {\bibinfo {volume} {107}},\ \bibinfo {pages}
  {126402} (\bibinfo {year} {2011})}\BibitemShut {NoStop}%
\bibitem [{\citenamefont {Feng}\ \emph {et~al.}(2014)\citenamefont {Feng},
  \citenamefont {Wang}, \citenamefont {Chen}, \citenamefont {Shi},
  \citenamefont {Xie}, \citenamefont {Yi}, \citenamefont {Liang}, \citenamefont
  {He}, \citenamefont {He}, \citenamefont {Peng}, \citenamefont {Liu},
  \citenamefont {Liu}, \citenamefont {Zhao}, \citenamefont {Liu}, \citenamefont
  {Dong}, \citenamefont {Zhang}, \citenamefont {Chen}, \citenamefont {Xu},
  \citenamefont {Dai}, \citenamefont {Fang},\ and\ \citenamefont
  {Zhou}}]{Feng2014}%
  \BibitemOpen
  \bibfield  {author} {\bibinfo {author} {\bibfnamefont {Y.}~\bibnamefont
  {Feng}}, \bibinfo {author} {\bibfnamefont {Z.}~\bibnamefont {Wang}}, \bibinfo
  {author} {\bibfnamefont {C.}~\bibnamefont {Chen}}, \bibinfo {author}
  {\bibfnamefont {Y.}~\bibnamefont {Shi}}, \bibinfo {author} {\bibfnamefont
  {Z.}~\bibnamefont {Xie}}, \bibinfo {author} {\bibfnamefont {H.}~\bibnamefont
  {Yi}}, \bibinfo {author} {\bibfnamefont {A.}~\bibnamefont {Liang}}, \bibinfo
  {author} {\bibfnamefont {S.}~\bibnamefont {He}}, \bibinfo {author}
  {\bibfnamefont {J.}~\bibnamefont {He}}, \bibinfo {author} {\bibfnamefont
  {Y.}~\bibnamefont {Peng}}, \bibinfo {author} {\bibfnamefont {X.}~\bibnamefont
  {Liu}}, \bibinfo {author} {\bibfnamefont {Y.}~\bibnamefont {Liu}}, \bibinfo
  {author} {\bibfnamefont {L.}~\bibnamefont {Zhao}}, \bibinfo {author}
  {\bibfnamefont {G.}~\bibnamefont {Liu}}, \bibinfo {author} {\bibfnamefont
  {X.}~\bibnamefont {Dong}}, \bibinfo {author} {\bibfnamefont {J.}~\bibnamefont
  {Zhang}}, \bibinfo {author} {\bibfnamefont {C.}~\bibnamefont {Chen}},
  \bibinfo {author} {\bibfnamefont {Z.}~\bibnamefont {Xu}}, \bibinfo {author}
  {\bibfnamefont {X.}~\bibnamefont {Dai}}, \bibinfo {author} {\bibfnamefont
  {Z.}~\bibnamefont {Fang}}, \ and\ \bibinfo {author} {\bibfnamefont {X.~J.}\
  \bibnamefont {Zhou}},\ }\href {http://dx.doi.org/10.1038/srep05385}
  {\bibfield  {journal} {\bibinfo  {journal} {Sci. Rep.}\ }\textbf {\bibinfo
  {volume} {4}},\ \bibinfo {pages} {5385} (\bibinfo {year} {2014})}\BibitemShut
  {NoStop}%
\bibitem [{\citenamefont {Guo}\ \emph {et~al.}(2014)\citenamefont {Guo},
  \citenamefont {Princep}, \citenamefont {Zhang}, \citenamefont {Manuel},
  \citenamefont {Khalyavin}, \citenamefont {Mazin}, \citenamefont {Shi},\ and\
  \citenamefont {Boothroyd}}]{Guo2014}%
  \BibitemOpen
  \bibfield  {author} {\bibinfo {author} {\bibfnamefont {Y.~F.}\ \bibnamefont
  {Guo}}, \bibinfo {author} {\bibfnamefont {A.~J.}\ \bibnamefont {Princep}},
  \bibinfo {author} {\bibfnamefont {X.}~\bibnamefont {Zhang}}, \bibinfo
  {author} {\bibfnamefont {P.}~\bibnamefont {Manuel}}, \bibinfo {author}
  {\bibfnamefont {D.}~\bibnamefont {Khalyavin}}, \bibinfo {author}
  {\bibfnamefont {I.~I.}\ \bibnamefont {Mazin}}, \bibinfo {author}
  {\bibfnamefont {Y.~G.}\ \bibnamefont {Shi}}, \ and\ \bibinfo {author}
  {\bibfnamefont {A.~T.}\ \bibnamefont {Boothroyd}},\ }\href {\doibase
  10.1103/PhysRevB.90.075120} {\bibfield  {journal} {\bibinfo  {journal} {Phys.
  Rev. B}\ }\textbf {\bibinfo {volume} {90}},\ \bibinfo {pages} {075120}
  (\bibinfo {year} {2014})}\BibitemShut {NoStop}%
\bibitem [{\citenamefont {Wang}\ \emph {et~al.}(2012)\citenamefont {Wang},
  \citenamefont {Graf}, \citenamefont {Wang}, \citenamefont {Lei},
  \citenamefont {Tozer},\ and\ \citenamefont {Petrovic}}]{Wang2012}%
  \BibitemOpen
  \bibfield  {author} {\bibinfo {author} {\bibfnamefont {K.}~\bibnamefont
  {Wang}}, \bibinfo {author} {\bibfnamefont {D.}~\bibnamefont {Graf}}, \bibinfo
  {author} {\bibfnamefont {L.}~\bibnamefont {Wang}}, \bibinfo {author}
  {\bibfnamefont {H.}~\bibnamefont {Lei}}, \bibinfo {author} {\bibfnamefont
  {S.~W.}\ \bibnamefont {Tozer}}, \ and\ \bibinfo {author} {\bibfnamefont
  {C.}~\bibnamefont {Petrovic}},\ }\href {\doibase 10.1103/PhysRevB.85.041101}
  {\bibfield  {journal} {\bibinfo  {journal} {Phys. Rev. B}\ }\textbf {\bibinfo
  {volume} {85}},\ \bibinfo {pages} {041101} (\bibinfo {year}
  {2012})}\BibitemShut {NoStop}%
\bibitem [{\citenamefont {He}\ \emph {et~al.}(2012)\citenamefont {He},
  \citenamefont {Wang},\ and\ \citenamefont {Chen}}]{He2012}%
  \BibitemOpen
  \bibfield  {author} {\bibinfo {author} {\bibfnamefont {J.~B.}\ \bibnamefont
  {He}}, \bibinfo {author} {\bibfnamefont {D.~M.}\ \bibnamefont {Wang}}, \ and\
  \bibinfo {author} {\bibfnamefont {G.~F.}\ \bibnamefont {Chen}},\ }\href
  {\doibase http://dx.doi.org/10.1063/1.3694760} {\bibfield  {journal}
  {\bibinfo  {journal} {Appl. Phys. Lett.}\ }\textbf {\bibinfo {volume}
  {100}},\ \bibinfo {eid} {112405} (\bibinfo {year} {2012})}\BibitemShut
  {NoStop}%
\bibitem [{sup()}]{supplemental}%
  \BibitemOpen
  \href@noop {} {\bibinfo  {journal} {{ See Supplemental Material attached to
  this preprint for the critical exponent analysis, magnetic susceptibility
  data, analytic expressions for the spin-wave gaps and magnon band- widths,
  procedure used to determine the best-fit spin-wave model, and detailed
  comparisons between the observed and calculated magnon spectra}}\
  }\BibitemShut {NoStop}%
\bibitem [{\citenamefont {Hiess}\ \emph {et~al.}(2006)\citenamefont {Hiess},
  \citenamefont {Jim\'{e}nez-Ruiz}, \citenamefont {Courtois}, \citenamefont
  {Currat}, \citenamefont {Kulda},\ and\ \citenamefont {Bermejo}}]{Hiess2006}%
  \BibitemOpen
\bibfield  {journal} {  }\bibfield  {author} {\bibinfo {author} {\bibfnamefont
  {A.}~\bibnamefont {Hiess}}, \bibinfo {author} {\bibfnamefont
  {M.}~\bibnamefont {Jim\'{e}nez-Ruiz}}, \bibinfo {author} {\bibfnamefont
  {P.}~\bibnamefont {Courtois}}, \bibinfo {author} {\bibfnamefont
  {R.}~\bibnamefont {Currat}}, \bibinfo {author} {\bibfnamefont
  {J.}~\bibnamefont {Kulda}}, \ and\ \bibinfo {author} {\bibfnamefont
  {F.}~\bibnamefont {Bermejo}},\ }\href {\doibase
  http://dx.doi.org/10.1016/j.physb.2006.05.370} {\bibfield  {journal}
  {\bibinfo  {journal} {Physica B}\ }\textbf {\bibinfo {volume} {385-386}},\
  \bibinfo {pages} {1077 } (\bibinfo {year} {2006})}\BibitemShut {NoStop}%
\bibitem [{\citenamefont {Kempa}\ \emph {et~al.}(2006)\citenamefont {Kempa},
  \citenamefont {Janousova}, \citenamefont {Saroun}, \citenamefont {Flores},
  \citenamefont {Boehm}, \citenamefont {Demmel},\ and\ \citenamefont
  {Kulda}}]{Kempa2006}%
  \BibitemOpen
  \bibfield  {author} {\bibinfo {author} {\bibfnamefont {M.}~\bibnamefont
  {Kempa}}, \bibinfo {author} {\bibfnamefont {B.}~\bibnamefont {Janousova}},
  \bibinfo {author} {\bibfnamefont {J.}~\bibnamefont {Saroun}}, \bibinfo
  {author} {\bibfnamefont {P.}~\bibnamefont {Flores}}, \bibinfo {author}
  {\bibfnamefont {M.}~\bibnamefont {Boehm}}, \bibinfo {author} {\bibfnamefont
  {F.}~\bibnamefont {Demmel}}, \ and\ \bibinfo {author} {\bibfnamefont
  {J.}~\bibnamefont {Kulda}},\ }\href {\doibase
  http://dx.doi.org/10.1016/j.physb.2006.05.371} {\bibfield  {journal}
  {\bibinfo  {journal} {Physica B}\ }\textbf {\bibinfo {volume} {385-386}},\
  \bibinfo {pages} {1080 } (\bibinfo {year} {2006})}\BibitemShut {NoStop}%
\bibitem [{\citenamefont {Singh}\ \emph {et~al.}(2009)\citenamefont {Singh},
  \citenamefont {Green}, \citenamefont {Huang}, \citenamefont {Kreyssig},
  \citenamefont {McQueeney}, \citenamefont {Johnston},\ and\ \citenamefont
  {Goldman}}]{Singh2009}%
  \BibitemOpen
  \bibfield  {author} {\bibinfo {author} {\bibfnamefont {Y.}~\bibnamefont
  {Singh}}, \bibinfo {author} {\bibfnamefont {M.~A.}\ \bibnamefont {Green}},
  \bibinfo {author} {\bibfnamefont {Q.}~\bibnamefont {Huang}}, \bibinfo
  {author} {\bibfnamefont {A.}~\bibnamefont {Kreyssig}}, \bibinfo {author}
  {\bibfnamefont {R.~J.}\ \bibnamefont {McQueeney}}, \bibinfo {author}
  {\bibfnamefont {D.~C.}\ \bibnamefont {Johnston}}, \ and\ \bibinfo {author}
  {\bibfnamefont {A.~I.}\ \bibnamefont {Goldman}},\ }\href {\doibase
  10.1103/PhysRevB.80.100403} {\bibfield  {journal} {\bibinfo  {journal} {Phys.
  Rev. B}\ }\textbf {\bibinfo {volume} {80}},\ \bibinfo {pages} {100403}
  (\bibinfo {year} {2009})}\BibitemShut {NoStop}%
\bibitem [{\citenamefont {Tegel}\ \emph {et~al.}(2008)\citenamefont {Tegel},
  \citenamefont {Rotter}, \citenamefont {Weiss}, \citenamefont {Schappacher},
  \citenamefont {Poettgen},\ and\ \citenamefont {Johrendt}}]{Tegel2008}%
  \BibitemOpen
  \bibfield  {author} {\bibinfo {author} {\bibfnamefont {M.}~\bibnamefont
  {Tegel}}, \bibinfo {author} {\bibfnamefont {M.}~\bibnamefont {Rotter}},
  \bibinfo {author} {\bibfnamefont {V.}~\bibnamefont {Weiss}}, \bibinfo
  {author} {\bibfnamefont {F.~M.}\ \bibnamefont {Schappacher}}, \bibinfo
  {author} {\bibfnamefont {R.}~\bibnamefont {Poettgen}}, \ and\ \bibinfo
  {author} {\bibfnamefont {D.}~\bibnamefont {Johrendt}},\ }\href
  {http://stacks.iop.org/0953-8984/20/i=45/a=452201} {\bibfield  {journal}
  {\bibinfo  {journal} {J. Phys.: Cond. Mat.}\ }\textbf {\bibinfo {volume}
  {20}},\ \bibinfo {pages} {452201} (\bibinfo {year} {2008})}\BibitemShut
  {NoStop}%
\bibitem [{\citenamefont {Wilson}\ \emph {et~al.}(2010)\citenamefont {Wilson},
  \citenamefont {Rotundu}, \citenamefont {Yamani}, \citenamefont {Valdivia},
  \citenamefont {Freelon}, \citenamefont {Bourret-Courchesne},\ and\
  \citenamefont {Birgeneau}}]{Wilson2010}%
  \BibitemOpen
  \bibfield  {author} {\bibinfo {author} {\bibfnamefont {S.~D.}\ \bibnamefont
  {Wilson}}, \bibinfo {author} {\bibfnamefont {C.~R.}\ \bibnamefont {Rotundu}},
  \bibinfo {author} {\bibfnamefont {Z.}~\bibnamefont {Yamani}}, \bibinfo
  {author} {\bibfnamefont {P.~N.}\ \bibnamefont {Valdivia}}, \bibinfo {author}
  {\bibfnamefont {B.}~\bibnamefont {Freelon}}, \bibinfo {author} {\bibfnamefont
  {E.}~\bibnamefont {Bourret-Courchesne}}, \ and\ \bibinfo {author}
  {\bibfnamefont {R.~J.}\ \bibnamefont {Birgeneau}},\ }\href {\doibase
  10.1103/PhysRevB.81.014501} {\bibfield  {journal} {\bibinfo  {journal} {Phys.
  Rev. B}\ }\textbf {\bibinfo {volume} {81}},\ \bibinfo {pages} {014501}
  (\bibinfo {year} {2010})}\BibitemShut {NoStop}%
\bibitem [{\citenamefont {Saroun}\ and\ \citenamefont
  {Kulda}(1997)}]{Saroun1997}%
  \BibitemOpen
  \bibfield  {author} {\bibinfo {author} {\bibfnamefont {J.}~\bibnamefont
  {Saroun}}\ and\ \bibinfo {author} {\bibfnamefont {J.}~\bibnamefont {Kulda}},\
  }\href {\doibase http://dx.doi.org/10.1016/S0921-4526(97)00037-9} {\bibfield
  {journal} {\bibinfo  {journal} {Physica B}\ }\textbf {\bibinfo {volume}
  {234}},\ \bibinfo {pages} {1102 } (\bibinfo {year} {1997})}\BibitemShut
  {NoStop}%
\bibitem [{\citenamefont {Saroun}\ and\ \citenamefont
  {Kulda}(2002)}]{Saroun2002}%
  \BibitemOpen
  \bibfield  {author} {\bibinfo {author} {\bibfnamefont {J.}~\bibnamefont
  {Saroun}}\ and\ \bibinfo {author} {\bibfnamefont {J.}~\bibnamefont {Kulda}},\
  }\href@noop {} {\bibfield  {journal} {\bibinfo  {journal} {Neutron News}\
  }\textbf {\bibinfo {volume} {13}},\ \bibinfo {pages} {15} (\bibinfo {year}
  {2002})}\BibitemShut {NoStop}%
\bibitem [{\citenamefont {Calder}\ \emph {et~al.}(2014)\citenamefont {Calder},
  \citenamefont {Saparov}, \citenamefont {Cao}, \citenamefont {Niedziela},
  \citenamefont {Lumsden}, \citenamefont {Sefat},\ and\ \citenamefont
  {Christianson}}]{Calder2014}%
  \BibitemOpen
  \bibfield  {author} {\bibinfo {author} {\bibfnamefont {S.}~\bibnamefont
  {Calder}}, \bibinfo {author} {\bibfnamefont {B.}~\bibnamefont {Saparov}},
  \bibinfo {author} {\bibfnamefont {H.~B.}\ \bibnamefont {Cao}}, \bibinfo
  {author} {\bibfnamefont {J.~L.}\ \bibnamefont {Niedziela}}, \bibinfo {author}
  {\bibfnamefont {M.~D.}\ \bibnamefont {Lumsden}}, \bibinfo {author}
  {\bibfnamefont {A.~S.}\ \bibnamefont {Sefat}}, \ and\ \bibinfo {author}
  {\bibfnamefont {A.~D.}\ \bibnamefont {Christianson}},\ }\href {\doibase
  10.1103/PhysRevB.89.064417} {\bibfield  {journal} {\bibinfo  {journal} {Phys.
  Rev. B}\ }\textbf {\bibinfo {volume} {89}},\ \bibinfo {pages} {064417}
  (\bibinfo {year} {2014})}\BibitemShut {NoStop}%
\bibitem [{\citenamefont {Johnston}\ \emph {et~al.}(2011)\citenamefont
  {Johnston}, \citenamefont {McQueeney}, \citenamefont {Lake}, \citenamefont
  {Honecker}, \citenamefont {Zhitomirsky}, \citenamefont {Nath}, \citenamefont
  {Furukawa}, \citenamefont {Antropov},\ and\ \citenamefont
  {Singh}}]{Johnston2011}%
  \BibitemOpen
  \bibfield  {author} {\bibinfo {author} {\bibfnamefont {D.~C.}\ \bibnamefont
  {Johnston}}, \bibinfo {author} {\bibfnamefont {R.~J.}\ \bibnamefont
  {McQueeney}}, \bibinfo {author} {\bibfnamefont {B.}~\bibnamefont {Lake}},
  \bibinfo {author} {\bibfnamefont {A.}~\bibnamefont {Honecker}}, \bibinfo
  {author} {\bibfnamefont {M.~E.}\ \bibnamefont {Zhitomirsky}}, \bibinfo
  {author} {\bibfnamefont {R.}~\bibnamefont {Nath}}, \bibinfo {author}
  {\bibfnamefont {Y.}~\bibnamefont {Furukawa}}, \bibinfo {author}
  {\bibfnamefont {V.~P.}\ \bibnamefont {Antropov}}, \ and\ \bibinfo {author}
  {\bibfnamefont {Y.}~\bibnamefont {Singh}},\ }\href {\doibase
  10.1103/PhysRevB.84.094445} {\bibfield  {journal} {\bibinfo  {journal} {Phys.
  Rev. B}\ }\textbf {\bibinfo {volume} {84}},\ \bibinfo {pages} {094445}
  (\bibinfo {year} {2011})}\BibitemShut {NoStop}%
\bibitem [{err()}]{erratum}%
  \BibitemOpen
  \href@noop {} {\bibinfo  {journal} {{Private communication, S. Calder,
  Quantum Condensed Matter Division, Oak Ridge National Laboratory, Oak Ridge,
  Tennessee 37831, USA}}\ }\BibitemShut {NoStop}%
\bibitem [{\citenamefont {Ewings}\ \emph {et~al.}(2011)\citenamefont {Ewings},
  \citenamefont {Perring}, \citenamefont {Gillett}, \citenamefont {Das},
  \citenamefont {Sebastian}, \citenamefont {Taylor}, \citenamefont {Guidi},\
  and\ \citenamefont {Boothroyd}}]{Ewings2011}%
  \BibitemOpen
\bibfield  {journal} {  }\bibfield  {author} {\bibinfo {author} {\bibfnamefont
  {R.~A.}\ \bibnamefont {Ewings}}, \bibinfo {author} {\bibfnamefont {T.~G.}\
  \bibnamefont {Perring}}, \bibinfo {author} {\bibfnamefont {J.}~\bibnamefont
  {Gillett}}, \bibinfo {author} {\bibfnamefont {S.~D.}\ \bibnamefont {Das}},
  \bibinfo {author} {\bibfnamefont {S.~E.}\ \bibnamefont {Sebastian}}, \bibinfo
  {author} {\bibfnamefont {A.~E.}\ \bibnamefont {Taylor}}, \bibinfo {author}
  {\bibfnamefont {T.}~\bibnamefont {Guidi}}, \ and\ \bibinfo {author}
  {\bibfnamefont {A.~T.}\ \bibnamefont {Boothroyd}},\ }\href@noop {} {\bibfield
   {journal} {\bibinfo  {journal} {Phys. Rev. B}\ }\textbf {\bibinfo {volume}
  {83}},\ \bibinfo {pages} {214519} (\bibinfo {year} {2011})}\BibitemShut
  {NoStop}%
\bibitem [{\citenamefont {Mambrini}\ \emph {et~al.}(2006)\citenamefont
  {Mambrini}, \citenamefont {L\"auchli}, \citenamefont {Poilblanc},\ and\
  \citenamefont {Mila}}]{Mambrini2006}%
  \BibitemOpen
  \bibfield  {author} {\bibinfo {author} {\bibfnamefont {M.}~\bibnamefont
  {Mambrini}}, \bibinfo {author} {\bibfnamefont {A.}~\bibnamefont {L\"auchli}},
  \bibinfo {author} {\bibfnamefont {D.}~\bibnamefont {Poilblanc}}, \ and\
  \bibinfo {author} {\bibfnamefont {F.}~\bibnamefont {Mila}},\ }\href {\doibase
  10.1103/PhysRevB.74.144422} {\bibfield  {journal} {\bibinfo  {journal} {Phys.
  Rev. B}\ }\textbf {\bibinfo {volume} {74}},\ \bibinfo {pages} {144422}
  (\bibinfo {year} {2006})}\BibitemShut {NoStop}%
\bibitem [{\citenamefont {Richter}\ and\ \citenamefont
  {Schulenburg}(2010)}]{Richter2010}%
  \BibitemOpen
  \bibfield  {author} {\bibinfo {author} {\bibfnamefont {J.}~\bibnamefont
  {Richter}}\ and\ \bibinfo {author} {\bibfnamefont {J.}~\bibnamefont
  {Schulenburg}},\ }\href {\doibase 10.1140/epjb/e2009-00400-4} {\bibfield
  {journal} {\bibinfo  {journal} {Eur. Phys. J. B}\ }\textbf {\bibinfo {volume}
  {73}},\ \bibinfo {pages} {117} (\bibinfo {year} {2010})}\BibitemShut
  {NoStop}%
\bibitem [{\citenamefont {Wang}(2012)}]{Wang2012a}%
  \BibitemOpen
  \bibfield  {author} {\bibinfo {author} {\bibfnamefont {H.-Y.}\ \bibnamefont
  {Wang}},\ }\href {\doibase 10.1103/PhysRevB.86.144411} {\bibfield  {journal}
  {\bibinfo  {journal} {Phys. Rev. B}\ }\textbf {\bibinfo {volume} {86}},\
  \bibinfo {pages} {144411} (\bibinfo {year} {2012})}\BibitemShut {NoStop}%
\bibitem [{\citenamefont {May}\ \emph {et~al.}(2014)\citenamefont {May},
  \citenamefont {McGuire},\ and\ \citenamefont {Sales}}]{May2014}%
  \BibitemOpen
  \bibfield  {author} {\bibinfo {author} {\bibfnamefont {A.~F.}\ \bibnamefont
  {May}}, \bibinfo {author} {\bibfnamefont {M.~A.}\ \bibnamefont {McGuire}}, \
  and\ \bibinfo {author} {\bibfnamefont {B.~C.}\ \bibnamefont {Sales}},\ }\href
  {\doibase 10.1103/PhysRevB.90.075109} {\bibfield  {journal} {\bibinfo
  {journal} {Phys. Rev. B}\ }\textbf {\bibinfo {volume} {90}},\ \bibinfo
  {pages} {075109} (\bibinfo {year} {2014})}\BibitemShut {NoStop}%
\bibitem [{\citenamefont {Masuda}\ \emph {et~al.}(2016)\citenamefont {Masuda},
  \citenamefont {Sakai}, \citenamefont {Tokunaga}, \citenamefont {Yamasaki},
  \citenamefont {Miyake}, \citenamefont {Shiogai}, \citenamefont {Nakamura},
  \citenamefont {Awaji}, \citenamefont {Tsukazaki}, \citenamefont {Nakao},
  \citenamefont {Murakami}, \citenamefont {Arima}, \citenamefont {Tokura},\
  and\ \citenamefont {Ishiwata}}]{Masuda2016}%
  \BibitemOpen
  \bibfield  {author} {\bibinfo {author} {\bibfnamefont {H.}~\bibnamefont
  {Masuda}}, \bibinfo {author} {\bibfnamefont {H.}~\bibnamefont {Sakai}},
  \bibinfo {author} {\bibfnamefont {M.}~\bibnamefont {Tokunaga}}, \bibinfo
  {author} {\bibfnamefont {Y.}~\bibnamefont {Yamasaki}}, \bibinfo {author}
  {\bibfnamefont {A.}~\bibnamefont {Miyake}}, \bibinfo {author} {\bibfnamefont
  {J.}~\bibnamefont {Shiogai}}, \bibinfo {author} {\bibfnamefont
  {S.}~\bibnamefont {Nakamura}}, \bibinfo {author} {\bibfnamefont
  {S.}~\bibnamefont {Awaji}}, \bibinfo {author} {\bibfnamefont
  {A.}~\bibnamefont {Tsukazaki}}, \bibinfo {author} {\bibfnamefont
  {H.}~\bibnamefont {Nakao}}, \bibinfo {author} {\bibfnamefont
  {Y.}~\bibnamefont {Murakami}}, \bibinfo {author} {\bibfnamefont {T.-H.}\
  \bibnamefont {Arima}}, \bibinfo {author} {\bibfnamefont {Y.}~\bibnamefont
  {Tokura}}, \ and\ \bibinfo {author} {\bibfnamefont {S.}~\bibnamefont
  {Ishiwata}},\ }\href@noop {} {\bibfield  {journal} {\bibinfo  {journal} {Sci.
  Adv.}\ }\textbf {\bibinfo {volume} {2}},\ \bibinfo {pages} {e1501117}
  (\bibinfo {year} {2016})}\BibitemShut {NoStop}%
\bibitem [{\citenamefont {Borisenko}\ \emph {et~al.}()\citenamefont
  {Borisenko}, \citenamefont {Evtushinsky}, \citenamefont {Gibson},
  \citenamefont {Yaresko}, \citenamefont {Kim}, \citenamefont {Ali},
  \citenamefont {Buechner}, \citenamefont {Hoesch},\ and\ \citenamefont
  {Cava}}]{Borisenko2016}%
  \BibitemOpen
  \bibfield  {author} {\bibinfo {author} {\bibfnamefont {S.}~\bibnamefont
  {Borisenko}}, \bibinfo {author} {\bibfnamefont {D.}~\bibnamefont
  {Evtushinsky}}, \bibinfo {author} {\bibfnamefont {Q.}~\bibnamefont {Gibson}},
  \bibinfo {author} {\bibfnamefont {A.}~\bibnamefont {Yaresko}}, \bibinfo
  {author} {\bibfnamefont {T.}~\bibnamefont {Kim}}, \bibinfo {author}
  {\bibfnamefont {M.~N.}\ \bibnamefont {Ali}}, \bibinfo {author} {\bibfnamefont
  {B.}~\bibnamefont {Buechner}}, \bibinfo {author} {\bibfnamefont
  {M.}~\bibnamefont {Hoesch}}, \ and\ \bibinfo {author} {\bibfnamefont {R.~J.}\
  \bibnamefont {Cava}},\ }\href@noop {} {\bibinfo  {journal}
  {arXiv:1507.04847}\ }\BibitemShut {NoStop}%
\bibitem [{\citenamefont {Zhang}\ \emph {et~al.}(2016)\citenamefont {Zhang},
  \citenamefont {Liu}, \citenamefont {Yi}, \citenamefont {Zhao}, \citenamefont
  {Xia}, \citenamefont {Ji}, \citenamefont {Shi}, \citenamefont {Yu},
  \citenamefont {Wang}, \citenamefont {Chen},\ and\ \citenamefont
  {Zhang}}]{ZhangLiuYiEtAl2016}%
  \BibitemOpen
\bibfield  {journal} {  }\bibfield  {author} {\bibinfo {author} {\bibfnamefont
  {A.}~\bibnamefont {Zhang}}, \bibinfo {author} {\bibfnamefont
  {C.}~\bibnamefont {Liu}}, \bibinfo {author} {\bibfnamefont {C.}~\bibnamefont
  {Yi}}, \bibinfo {author} {\bibfnamefont {G.}~\bibnamefont {Zhao}}, \bibinfo
  {author} {\bibfnamefont {T.-l.}\ \bibnamefont {Xia}}, \bibinfo {author}
  {\bibfnamefont {J.}~\bibnamefont {Ji}}, \bibinfo {author} {\bibfnamefont
  {Y.}~\bibnamefont {Shi}}, \bibinfo {author} {\bibfnamefont {R.}~\bibnamefont
  {Yu}}, \bibinfo {author} {\bibfnamefont {X.}~\bibnamefont {Wang}}, \bibinfo
  {author} {\bibfnamefont {C.}~\bibnamefont {Chen}}, \ and\ \bibinfo {author}
  {\bibfnamefont {Q.}~\bibnamefont {Zhang}},\ }\href
  {http://www.ncbi.nlm.nih.gov/pmc/articles/PMC5172363/} {\bibfield  {journal}
  {\bibinfo  {journal} {Nat. Commun.}\ }\textbf {\bibinfo {volume} {7}},\
  \bibinfo {pages} {13833} (\bibinfo {year} {2016})}\BibitemShut {NoStop}%
\end{thebibliography}%

\newpage
\begin{figure*}
\includegraphics[width=1.9\columnwidth,trim= 0pt 0pt 0pt 0pt, clip]{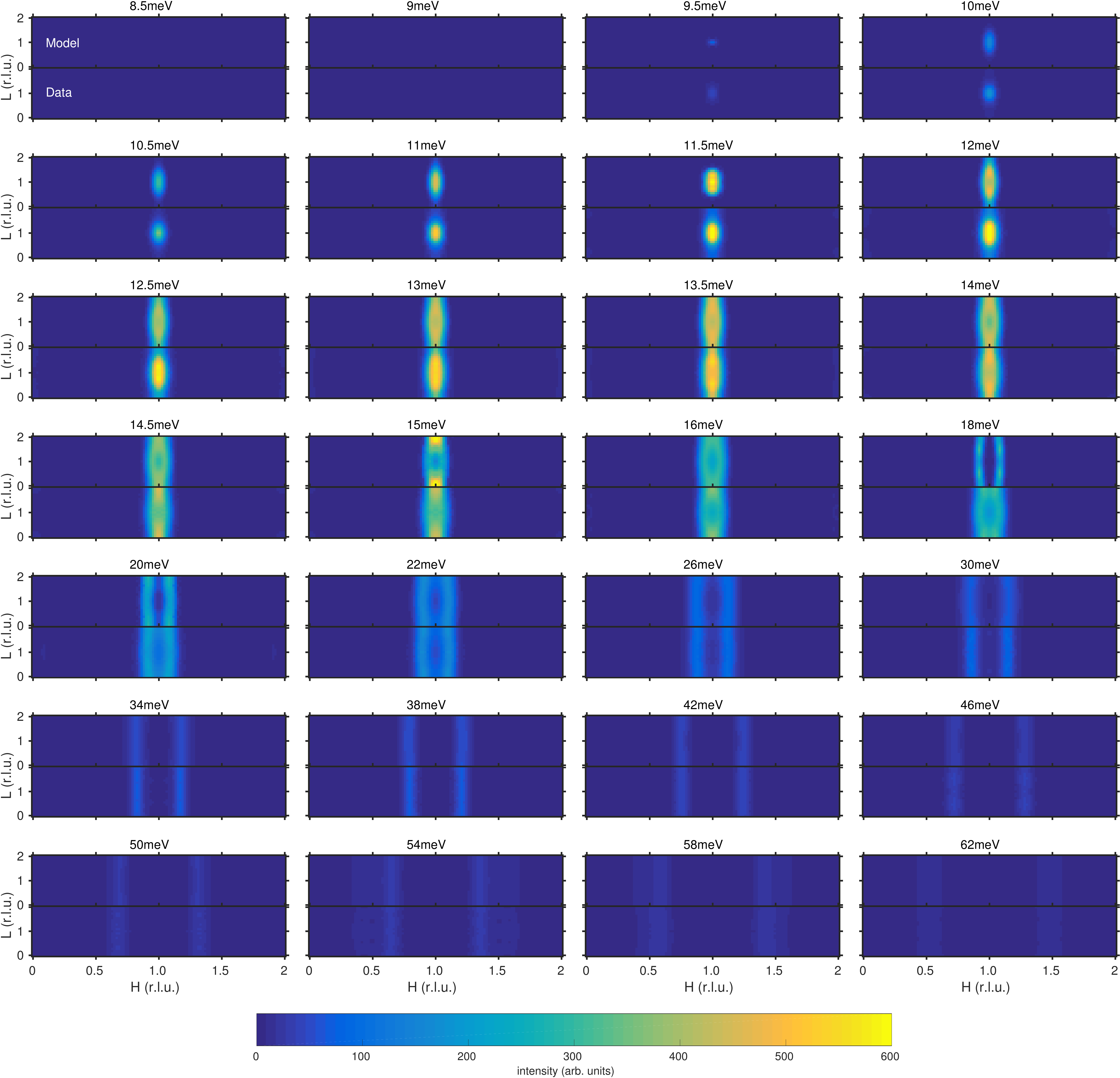}%
\\[5pt]
\justify{\label{sfig2} Fig.~S4 (color online). SrMnBi$_2$: Constant-energy maps of a 2$\times$2 r.l.u. area of the $(H0L)$ plane of reciprocal space. Each double panel shows the processed data (lower panel, explanation see text) as well as the best fit with a phenomenologial gaussian broadening (upper panel).}
\end{figure*}

\begin{figure*}
\includegraphics[width=1.9\columnwidth,trim= 0pt 0pt 0pt 0pt, clip]{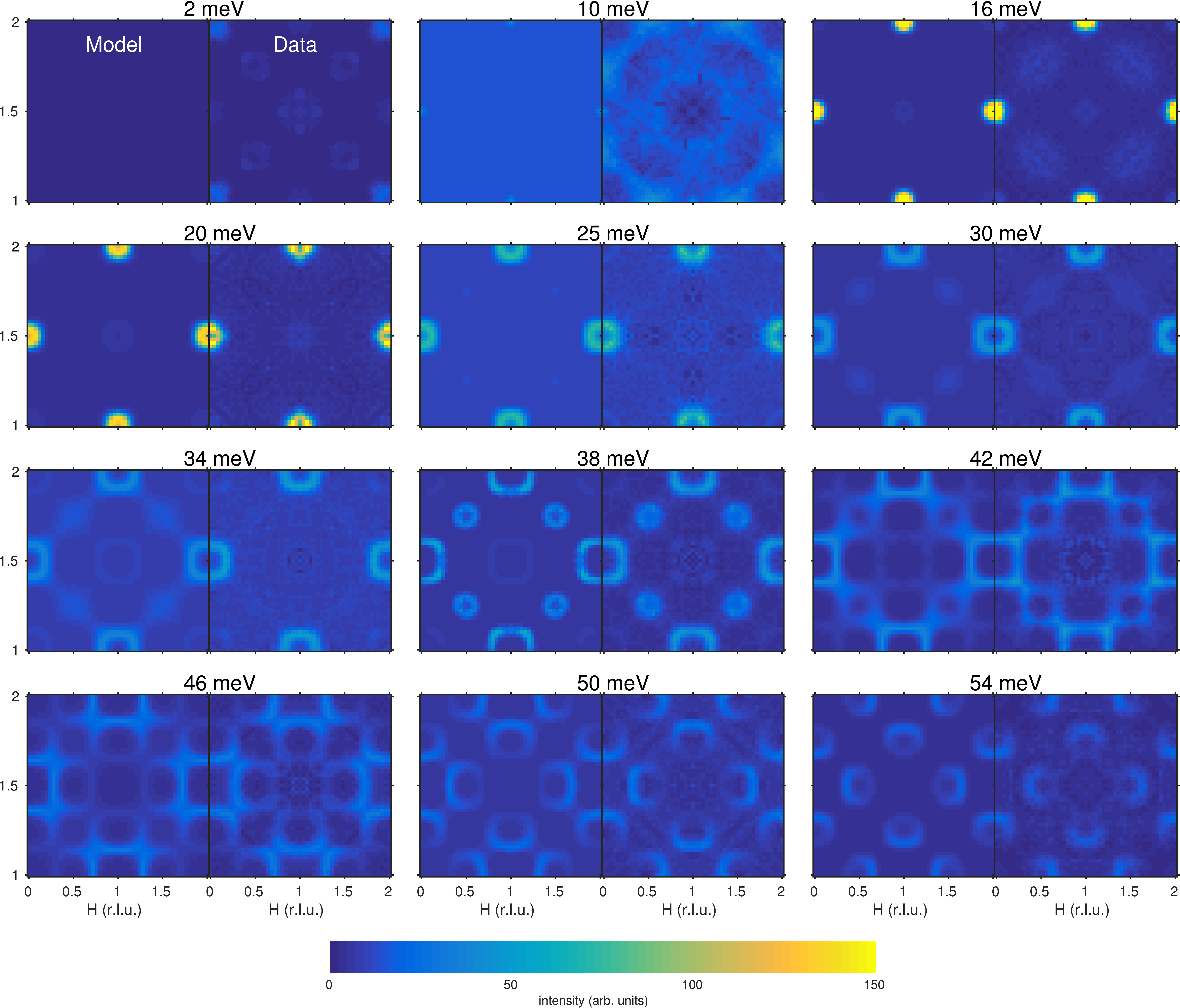}%
\\[5pt]
\justify{\label{sfig3} Fig.~S5 (color online). SrMnBi$_2$: Constant-energy maps of a 2$\times$2 r.l.u. area of the $(HK0)$ plane of reciprocal space. Each double panel shows the processed data (right panel, explanation see text) as well as the best fit with a phenomenologial gaussian broadening (left panel).}
\end{figure*}

\begin{figure*}
\includegraphics[width=1.9\columnwidth,trim= 0pt 0pt 0pt 0pt, clip]{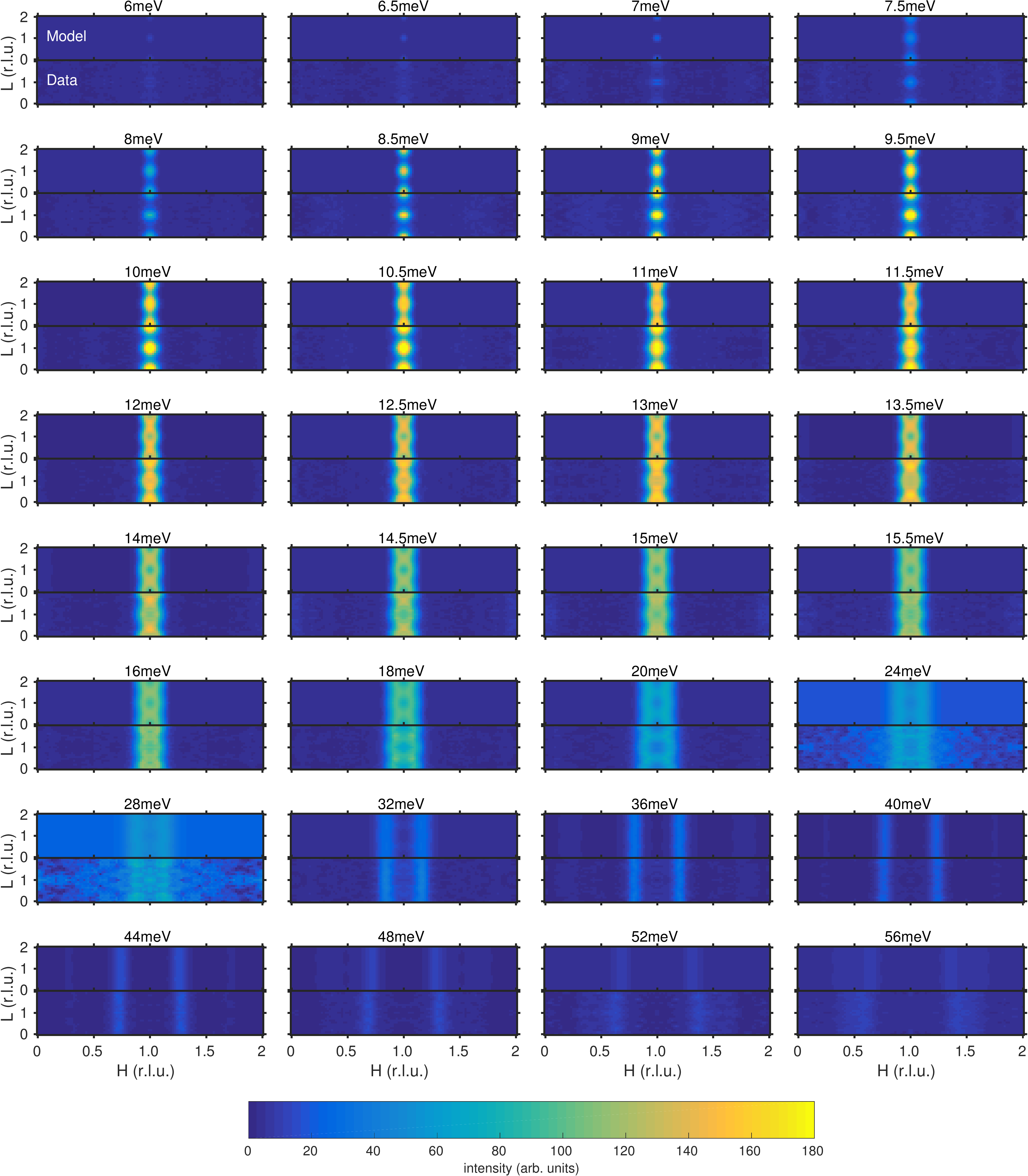}%
\\[5pt]
\justify{\label{sfig4} Fig.~S6 (color online). CaMnBi$_2$: Constant-energy maps of a 2$\times$2 r.l.u. area of the $(H0L)$ plane of reciprocal space. Each double panel shows the processed data (lower panel, explanation see text) as well as the best fit with a phenomenologial gaussian broadening (upper panel).}
\end{figure*}

\begin{figure*}
\includegraphics[width=1.9\columnwidth,trim= 0pt 0pt 0pt 0pt, clip]{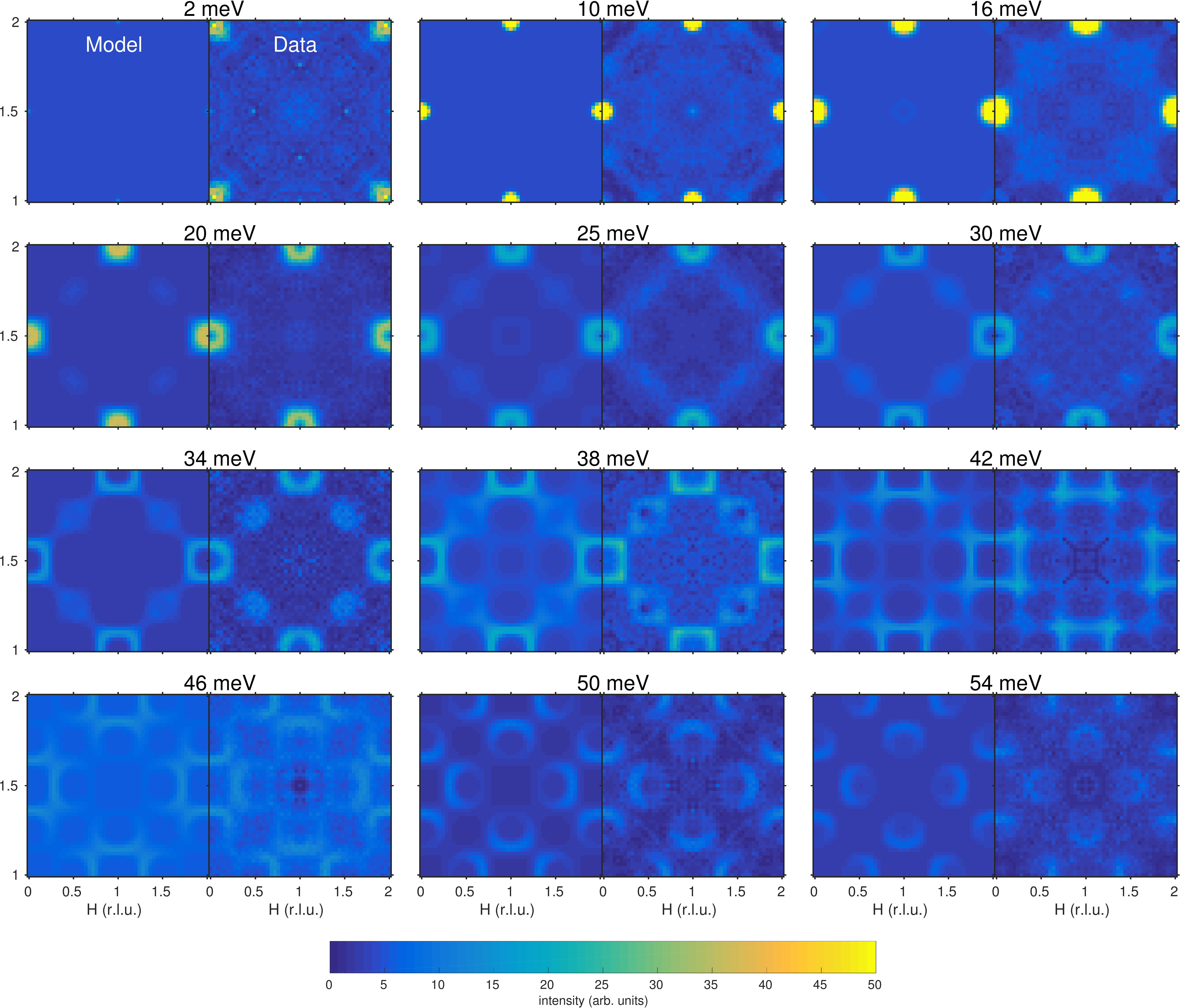}%
\\[5pt]
\justify{\label{sfig5} Fig.~S7 (color online). CaMnBi$_2$: Constant-energy maps of a 2$\times$2 r.l.u. area of the $(HK0)$ plane of reciprocal space. Each double panel shows the processed data (right panel, explanation see text) as well as the best fit with a phenomenologial gaussian broadening (left panel).}
\end{figure*}

\begin{figure*}
\includegraphics[width=1.8\columnwidth,trim= 0pt 0pt 0pt 0pt, clip]{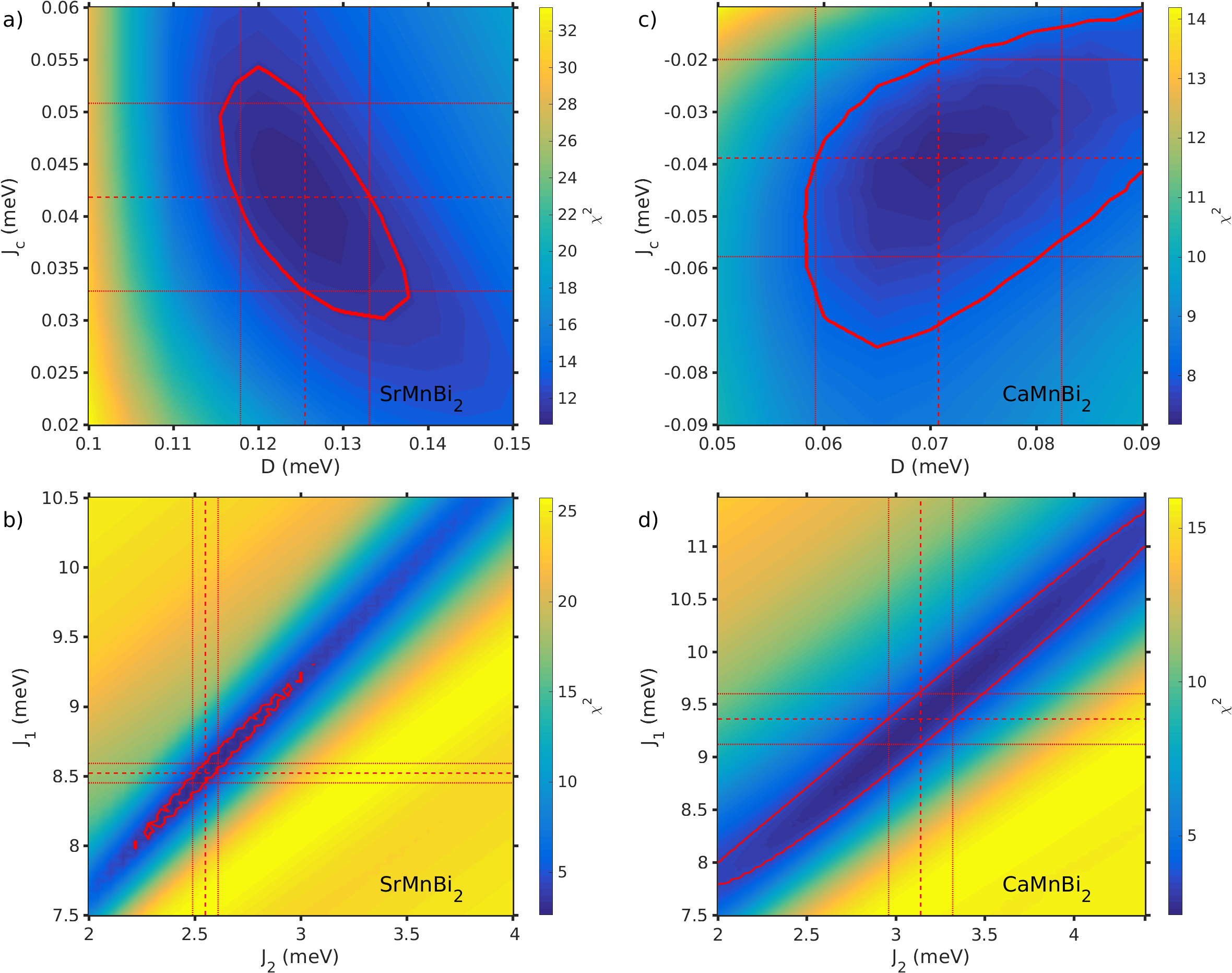}%
\\[5pt]
\justify{\label{sfig1} Fig.~S8 (color online). $\chi^2$ maps of least-squares fits of the linear spin-wave model for SrMnBi$_2$ (a,b) and CaMnBi$_2$ (c,d). Panels (a) and (c) reflect fits of a low energy ($E\leq20\,$\,meV) cut along the $(10L)$ direction of reciprocal space, using the resolution properties calculated by the RESTRAX ray-tracing routine \cite{Saroun1997,Saroun2002} (best-fit result shown in Fig.~6 of the main article). Panels (b) and (d) illustrate fit results of the high-energy ($30\leq E\leq 45\,$meV) dispersion in the $(HK0)$ plane, used to determine $J_1$ and $J_2$. The red lines represent the one-$\sigma$ contour. Dashed lines indicate the best-fit values and error margins quoted in the main article. Note that while $J_1$ and $J_2$ appear to be strongly correlated, weak details of the dispersion in the $(HK0)$ plane are not adequately reflected in the $\chi^2$ value and constrain the values further than suggested by the one-$\sigma$ contour.}
\end{figure*}

\end{document}